\documentclass[longauth, twocolumn]{aa}

\usepackage[T1]{fontenc}
\usepackage{ae,aecompl}
\usepackage{natbib}
\usepackage{graphicx}
\usepackage{txfonts}
\usepackage{array}
\usepackage{subfigure} %Sous-figures
\usepackage[skip=5pt]{caption} %Legendes
\usepackage{rotating}
\usepackage{amssymb}
\usepackage{multirow}

\usepackage{amsmath}

\def\msol{\hbox{\kern 0.20em $M_\odot$}}
\def\lsol{\hbox{\kern 0.20em $L_\odot$}}
\def\rsol{\hbox{\kern 0.20em $R_\odot$}}
\def\sr{\hbox{\kern 0.20em sr}}
\def\srmu{\hbox{\kern 0.20em sr$^{-1}$}}
\def\g{\hbox{\kern 0.20em g}}
\def\gmu{\hbox{\kern 0.20em g$^{-1}$}}
\def\kg{\hbox{\kern 0.20em kg}}
\def\pc{\hbox{\kern 0.20em pc}}
\def\mum{\hbox{\kern 0.20em $\mu$m}}
\def\mumd{\hbox{\kern 0.20em $\mu$m$^{-2}$}}
\def\cm{\hbox{\kern 0.20em cm}}
\def\m{\hbox{\kern 0.20em m}}
\def\km{\hbox{\kern 0.20em km}}
\def\nm{\hbox{\kern 0.20em nm}}
\def\s{\hbox{\kern 0.20em s}}
\def\h{\hbox{\kern 0.20em h}}
\def\sec{\hbox{\kern 0.20em sec}}
\def\min{\hbox {\kern 0.20em min}}
\def\smu{\hbox{\kern 0.20em s$^{-1}$}}
\def\smd{\hbox{\kern 0.20em s$^{-2}$}}
\def\an{\hbox{\kern 0.20em an}}
\def\anmu{\hbox{\kern 0.20em an$^{-1}$}}
\def\deg{\hbox{\kern 0.20em $^{\rm o}$}}
\def\yr{\hbox{\kern 0.20em yr}}
\def\yrmu{\hbox{\kern 0.20em yr$^{-1}$}}
\def\Myr{\hbox{\kern 0.20em Myr}}
\def\Mymu{\hbox{\kern 0.20em Myr$^{-1}$}}
\def\K{\hbox{\kern 0.20em K}}
\def\pcmu{\hbox{\kern 0.20em pc$^{-1}$}}
\def\pcmd{\hbox{\kern 0.20em pc$^{-2}$}}
\def\pcmt{\hbox{\kern 0.20em pc$^{-3}$}}
\def\kms{\hbox{\kern 0.20em km\kern 0.20em s$^{-1}$}}
\def\kmpd{\hbox{\kern 0.20em km$^{2}$}}
\def\kpc{\hbox{\kern 0.20em kpc}}
\def\cms{\hbox{\kern 0.20em cm\kern 0.20em s$^{-1}$}}
\def\erg{\hbox{\kern 0.20em erg}}
\def\ergs{\hbox{\kern 0.20em erg}}
\def\cmpd{\hbox{\kern 0.20em cm$^2$}}
\def\cmmd{\hbox{\kern 0.20em cm$^{-2}$}}
\def\cmms{\hbox{\kern 0.20em cm$^{-6}$}}
\def\cmpt{\hbox{\kern 0.20em cm$^3$}}
\def\cmmt{\hbox{\kern 0.20em cm$^{-3}$}}
\def\mpd{\hbox{\kern 0.20em m$^2$}}
\def\mmd{\hbox{\kern 0.20em m$^{-2}$}}
\def\mpt{\hbox{\kern 0.20em m$^3$}}
\def\mmt{\hbox{\kern 0.20em m$^{-3}$}}
\def\mujy{\hbox{\kern 0.20em $\mu$Jy}}
\def\mjy{\hbox{\kern 0.20em mJy}}
\def\Mj{\hbox{\kern 0.20em MJy}}
\def\jy{\hbox{\kern 0.20em Jy}}
\def\ghz{\hbox{\kern 0.20em GHz}}
\def\srmd{\hbox{\kern 0.20em sr$^{-1}$}}

\def \mum{$\mu$m}
\def\G{\hbox{\kern 0.20em G}}

\def\htwo{\hbox{H${}_2$}}
\def\h13cop{\hbox{H$^{13}$CO$^{+}$}}

\def\h2o{\hbox{H$_2$O}}

\title{Mass ejection and time variability in protostellar outflows: Cep\,E. SOLIS XVI}
\titlerunning{Mass ejection and time variability in Cep\,E}
\authorrunning{Schutzer et al.}

\usepackage[switch]{lineno}

\begin{document} 
%\linenumbers

\author{A. de A. Schutzer
          \inst{1}
         \and
          P. R. Rivera-Ortiz\inst{1}
          \and
          B. Lefloch\inst{1}
\and 
A. Gusdorf\inst{2,3}
\and
      C. Favre\inst{1}
\and
      D.~Segura-Cox\inst{4,5}\thanks{NSF Astronomy and Astrophysics Postdoctoral Fellow}
\and  
        A.~López-Sepulcre\inst{1,6}
\and 
        R.~Neri\inst{6}
\and
        J.~Ospina-Zamudio\inst{1}
\and 
        M.~De~Simone\inst{1}
\and
        C.~Codella\inst{7,1}
\and
        S.~Viti\inst{8,9}
\and 
        L.~Podio\inst{7}
\and
        J.~Pineda\inst{4}
\and
        R.~O'Donoghue\inst{9}
\and
        C.~Ceccarelli\inst{1}
\and
        P.~Caselli\inst{4}
\and F.~Alves\inst{4}
\and R.~Bachiller\inst{10}
\and N.~Balucani\inst{11,1,7}
\and E.~Bianchi\inst{1}
\and L.~Bizzocchi\inst{4,12}
\and S.~Bottinelli\inst{13,14}
\and E.~Caux\inst{13}
\and A.~Chac\'on-Tanarro\inst{10}
\and F.~Dulieu\inst{15}
\and J.~Enrique-Romero\inst{1,16}
\and F.~Fontani\inst{7}
\and S.~Feng\inst{4}
\and J.~Holdship\inst{8,9}
\and I.~Jim\'enez-Serra\inst{17}
\and A.~Jaber Al-Edhari\inst{1,18}
\and C.~Kahane\inst{1}
\and V.~Lattanzi\inst{4}
\and Y.~Oya\inst{19}
\and A.~Punanova\inst{20}
\and A.~Rimola\inst{16}
\and N.~Sakai\inst{21}
\and S.~Spezzano\inst{4}
\and I.~R.~Sims\inst{22}
\and V.~Taquet\inst{7}
\and L.~Testi\inst{23,7}
\and P.~Theul\'e\inst{24}
\and P.~Ugliengo\inst{25}
\and C.~Vastel\inst{13,14}
\and A.~I.~Vasyunin\inst{20}
\and F.~Vazart\inst{1}
\and S.~Yamamoto\inst{19}
\and A.~Witzel\inst{1}
 }
\institute{
%1
   Univ. Grenoble Alpes, CNRS, IPAG, 38000 Grenoble, France
              \email{andre.schutzer@univ-grenoble-alpes.fr}
\and
%2
    Laboratoire de Physique de l’ENS, ENS, Université PSL, CNRS, Sorbonne Université, Université de Paris, 75005, Paris, France 
    \and 
    %3
    Observatoire de Paris, PSL University, Sorbonne Université, LERMA, 75014, Paris, France
    \and
    %4
    Max-Planck-Institut f\"ur extraterrestrische Physik, Giessenbachstrasse 1, 85748 Garching, Germany       
    \and
    %5
    Department of Astronomy, The University of Texas at Austin, 2500 Speedway, Austin, TX, 78712, USA      
    \and 
    %6
    Institut de Radioastronomie Millimétrique (IRAM), 300 rue de la Piscine, 38406 Saint-Martin-D’Hères, France
    \and
%7
INAF, Osservatorio Astrofisico di Arcetri, Largo E. Fermi 5,50125 Firenze, Italy
\and 
%8
Leiden Observatory, Leiden University, PO Box 9513, 2300 RA Leiden, The Netherlands
\and
%9
Department of Physics and Astronomy, University College London, Gower Street, London, WC1E 6BT, UK
\and
%10
IGN, Observatorio Astron\'omico Nacional, Calle Alfonso XII, 28004 Madrid, Spain
\and
%11
Dipartimento di Chimica, Biologia e Biotecnologie, Via Elce di
Sotto 8, 06123 Perugia, Italy
\and
%12
Dipartimento di Chimica "G. Ciamician"  via F. Selmi 2, I-40126 Bologna, Italy
\and
%13
Universit\'e  de Toulouse, UPS-OMP, IRAP, Toulouse, France
\and
%14
CNRS, IRAP, 9 Av. Colonel Roche, BP 44346, 31028 Toulouse Cedex 4, France
\and 
%15
LERMA, Universit\'e de Cergy-Pontoise, Observatoire de Paris, 
PSL Research University, CNRS, Sorbonne Universit\'e, UPMC, Univ. Paris 06, 95000 Cergy Pontoise, France
\and 
%16
Departament de Qu\'{\i}mica, Universitat Aut\`onoma de Barcelona, 08193 Bellaterra, Catalonia, Spain
\and
%17
Centro de Astrobiología (CSIC, INTA), Ctra. de Ajalvir, km. 4,Torrejón de Ardoz, 28850 Madrid, Spain
\and
%18 
University of AL-Muthanna, College of Science, Physics Department, AL-Muthanna, Iraq
%School of Physics and Astronomy Queen Mary, University of London, 327 Mile %End Road, London, E1 4NS
\and
%19
Department of Physics, The University of Tokyo, 7-3-1, 
Hongo, Bunkyo-ku, Tokyo 113-0033, Japan
\and
%20
Ural Federal University, 620002, 19 Mira street, Yekaterinburg, Russia
\and
%21
The Institute of Physical and Chemical Research (RIKEN), 2-1, Hirosawa, 
Wako-shi, Saitama 351-0198, Japan
\and
%22
Univ Rennes, CNRS, IPR (Institut de Physique de Rennes) - UMR 6251, F-35000 Rennes, France
%
%Dipartimento di Fisica e Astronomia, Universit\`a degli Studi di Firenze, %Italy
\and
%23
ESO, Karl Schwarzchild Srt. 2, 85478 Garching bei M\"unchen, Germany
\and
%24
Aix-Marseille Universit\'e, PIIM UMR-CNRS 7345, 13397 Marseille, France
\and
%25
Universit\`a degli Studi di Torino, Dipartimento Chimica Via Pietro Giuria 7, 10125 Torino, Italy
}
   \date{Received 16 December 2021; accepted 08 March 2022}

\abstract{Protostellar jets are an important agent of star formation feedback, tightly connected with the mass-accretion process.  The history of jet formation and mass-ejection provides constraints  on the mass accretion history and the nature of the driving source.}{We want to characterize the time-variability of the mass-ejection phenomena at work in the Class 0 protostellar phase, in order to better understand the dynamics of the outflowing gas and bring more constraints on the origin of the jet chemical composition and the mass-accretion history.}{Using the NOEMA (NOrthern Extended Millimeter Array) interferometer, we have observed the emission of the CO 2--1 and SO $N_J$= $5_4$--$4_3$ rotational transitions at an angular resolution of $1.0\arcsec$ (820~au) and $0.4\arcsec$ (330~au), respectively,  towards the intermediate-mass Class 0 protostellar system Cep\,E.}{
The CO high-velocity jet emission reveals a central component of $\leq 400$~au diameter associated with high-velocity molecular knots, also detected in SO, surrounded by a collimated layer of entrained gas. The gas layer appears to accelerate along the main axis over a length scale $\delta_0\approx$ 700~au, while its diameter gradually increases up to several 1000~au at 2000~au from the protostar. The jet is fragmented into 18 knots of mass $\sim 10^{-3}\msol$, unevenly distributed between the northern and southern lobes, with velocity variations up to $15\kms$ close to the protostar, well below the jet terminal velocities in the northern ($+65\kms$) and southern ($-125\kms$) lobes, respectively. The knot interval distribution is approximately bi-modal with a  time scale of $\sim 50-80\yr$ close to the jet driving protostar Cep\,E-A and $\sim 150-200\yr$ at larger distances $> 12\arcsec$. The mass-loss rates derived from knot masses are overall steady, with values of $2.7\times 10^{-5}\msol\yrmu$ and $8.9\times 10^{-6}\msol\yrmu$ in the northern and southern lobe, respectively.} 
{The interaction of the ambient protostellar material with  high-velocity knots drives the formation of a molecular layer around the jet, which accounts for the higher mass-loss rate in the northern lobe. The jet dynamics are well accounted for by a simple precession model with a period of $2000\yr$ and a mass-ejection period of $55\yr$.}

\maketitle

\section{Introduction}

The earliest stages of the low-mass star formation process are associated with powerful mass-loss phenomena under the form of high-velocity ($\sim 100\kms$) collimated jets. These jets accelerate the ambient circumstellar gas and drive the formation of low-velocity ($\sim 10\kms$) molecular outflows \citep{Raga-Cabrit1993}, commonly observed in the millimeter lines of CO \citep{Bachiller1996}. Although the actual jet launch mechanism is still debated \citep[e.g.,][]{Bally2016}, it is well established that mass-loss phenomena take their origin in the inner 1--100~au around the nascent star and  their interaction with the dense surrounding protostellar gas is expected to play a major role in the formation of protostellar jets themselves \citep{Frank2014}. Jet/outflow systems have also been proposed to remove angular momentum from the central protostellar regions so to allow material accretion from the disk onto the central object \citep{Frank2014}. Therefore, the formation of jet/outflow systems  seems to be an indispensable mechanism in the process of mass accretion by the protostar. Retrieving the history of the jet may provide some constraints to the mass accretion history.

Time variability of the mass-ejection phenomena associated with young stellar objects was identified a long time ago from studies of Herbig-Haro (HH) objects, which revealed the presence of knots and wiggling structures, and from measurements of their proper motion, which showed direct evidence for velocity variations \citep{Reipurth-Bally-2001}. Close to the powering source(s) of the HH jet(s), chains of aligned knots are detected while at larger distances disconnected heads with lower velocities trace the impact against the ambient cloud. \citet{Raga1990} showed that velocity variability in the ejection process drives the formation of internal working surfaces, where the fast jet material catches up with the slow material, resulting into bright emission knots along the jet while the ejection direction variability produces a garden hose effect where the ejected material diverges from an equilibrium axis position. 
Distribution of knot dynamical ages measured in Class I atomic jets suggests that there are up to three superposed ejection modes with typical timescales of a few $10\yr$, $10^2\yr$ and $10^3\yr$, respectively \citep{Raga2002}.

The wiggling structure observed in molecular outflows has been interpreted and successfully modeled in terms of variability of the ejection direction \citep[see e.g.,][]{Gueth1996,Eisloeffel1996,Ferrero2015a,Podio2016,Lefevre2017}. Several 
theoretical models have been proposed to explain this phenomenon \citep{Masciadri-Raga2002,Terquem1999,Frank2014} but direct observational constraints are still lacking. It seems however that most of the identified precessing jets arise from multiple protostellar systems, which suggests that the presence of companion(s) probably plays a role in the origin of the phenomenon. The precession periods derived towards Class 0/I objects range from $\sim 400\yr$ to 5$\times 10^4\yr$ \citep{Frank2014}.

Recently, the Large Program CALYPSO (Continuum And Lines in Young ProtoStellar Objects) has significantly enhanced the statistics on the properties of molecular protostellar jets and their outflows by targeting a sample of 30 nearby ($d < 450\pc$) protostellar sources, mainly in the Class 0 stage, which were observed in selected millimeter transitions of CO, SiO and SO \citep{Podio2021}. The authors reported a high detection rate ($ > 80\%$) of high-velocity collimated jets for sources with $L > 1\lsol$, with a velocity asymmetry between the jet lobes for $\sim 30\%$ of the sample. Interestingly, half of the 12 protostellar jets detected in SiO display evidence for bending axes, suggestive of precession or wiggling. 

High-angular resolution observations of Class 0/I protostellar jets with ALMA and NOEMA, like e.g. HH111, HH212, IRAS04166+ 2706, L1157 \citep{Lefloch2007,Codella2007,Codella2014,Tafalla2010,Podio2016,Plunkett2015} bring strong evidence of time variability of the ejection process in the early protostellar phase. 
A few observational estimates on timescales  have been obtained, as illustrated by \citet{Plunkett2015}, who report an average value of $310\pm 150\yr$ between subsequent ejection events in outflow C7 in the cluster Serpens South, with a large scatter in the knot distribution. 
Also, in their study of a cluster of 46 outflow lobes in the  star forming region W43-MM1 at $5.5\kpc$, \cite{Nony2020} report clear events of episodic ejection. The typical timescale between 2 ejecta is estimated $\approx 500^{+300}_{-100}\yr$, consistent with the values reported in nearby low-mass protostars. Uncertainties remain high because of the poorly constrained geometry and the outflow inclination angle with respect to the line of sight. Moreover, as noticed by the authors, outflows observed at increasing angular resolution turn out to present several spatial (thus temporal) characteristic scales. Therefore, more observational work is needed in order to make progress on these questions. 

In this article, we present a detailed study of the high-velocity molecular jet from the young Class 0 intermediate-mass protostellar system CepE-mm \citep{Lefloch1996,Lefloch2015,Ospina-Zamudio2018}. Previous studies showed evidence for time-variability in the molecular outflow/jet emission, associated both with precession \citep{Eisloeffel1996} and internal bullets \citep{Gomez-Ruiz2012,Lefloch2015,Gusdorf2017}. Using the IRAM interferometer, we have imaged the rotational transitions CO $J$= 2--1 230538.00~MHz, a typical tracer of molecular outflows, along the full high-velocity jet at $1\arcsec$ resolution, and SO $N_J$= $5_4$--$4_3$ 206176.013~MHz, a probe of molecular jets, at $0.4\arcsec$ resolution towards the central protostellar core, as part of the SOLIS Large Program \citep{Ceccarelli2017}. We have carried out an observational study of the dynamics of the molecular outflow/jet system and the bullets. We show how the main jet kinematical features can be accounted for by a simple analytic model of the mass-ejection process. In a forthcoming paper (Schutzer {\em in prep}), we will present a detailed study of the unusually rich jet chemical composition and we will show how it originates from the jet structure and its dynamics.  

The paper is organized as follows. Section~2  summarizes the main properties of the Cep\,E-mm protostellar source. We present in Sect.~3 our interferometric observations of the outflow/jet in the CO 2--1 and SO $N_J$= $5_4$--$4_3$ lines. In Sect.~4, we analyze the kinematics of the CO jet/outflow, and we show that the observed jet acceleration actually traces the jet interaction  with the parental envelope. In Sect.~5, we present a detailed study of the molecular knots detected along the jet, and by means of a simple ballistic model we  constrain their formation mechanism and their physical properties (age, mass, dynamical age). In Sect.~6, we propose a new jet precession model, which aims to account for the outflow morphology.  Our conclusions are summarized in Sect.~7.

\section{The source}
\label{sec:source}

\begin{figure}[!th]
\begin{center}
\includegraphics[width=0.9\columnwidth]{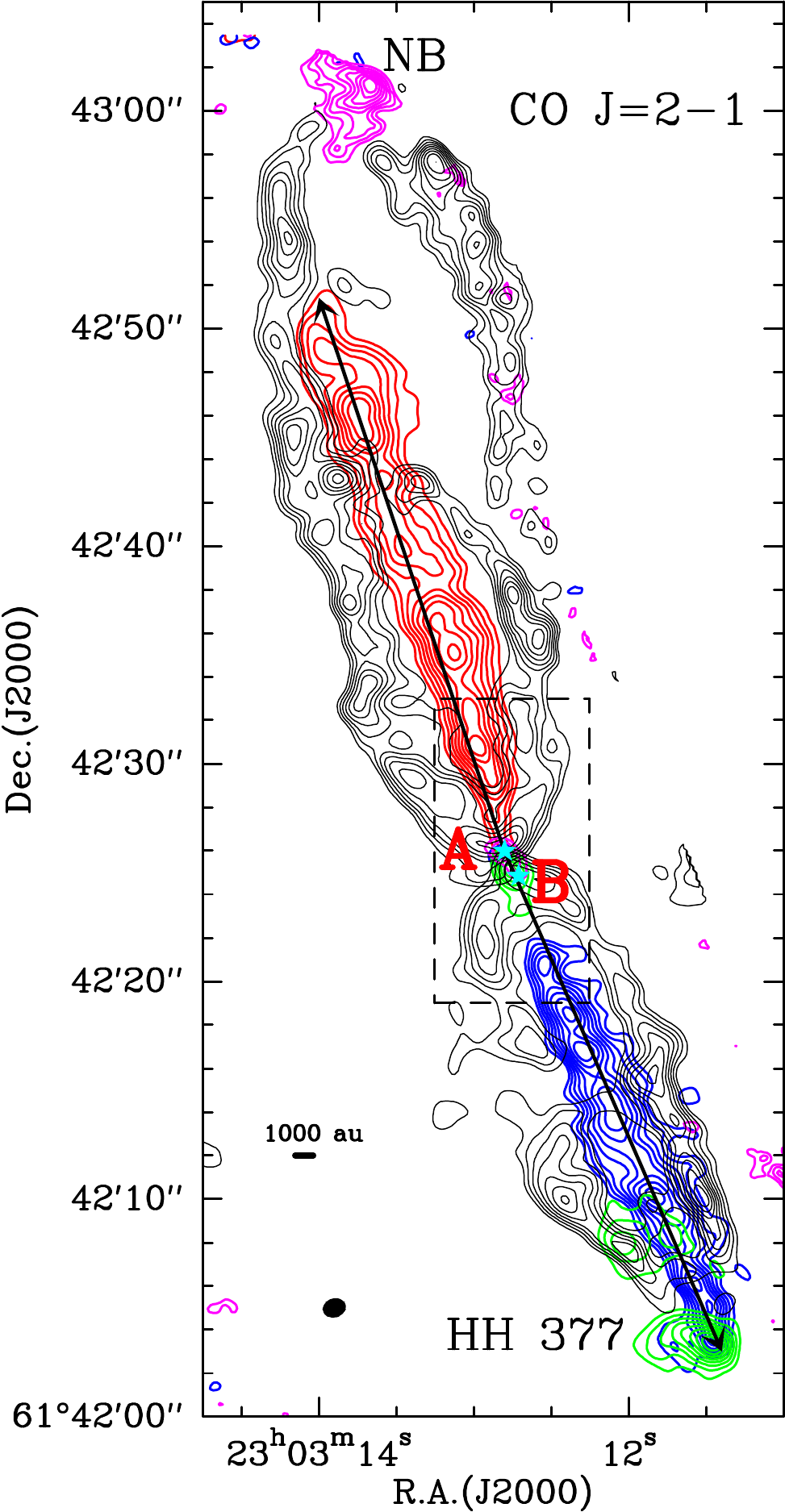} 
\caption{Emission of the Cep\,E-mm outflow as observed in the CO $2$--$1$ line with the PdBI at $1\arcsec$ resolution by \citet{Lefloch2015}.
Four main velocity components are detected: a)~the outflow cavity walls (black) emitting at low velocities, in the range [$-8$;$-3$]$\kms$ and [$-19$;$-14$]$\kms$ in the northern and southern lobe, respectively; b)~the jet, emitting at high velocities in the  range [$-135$;$-110$]$\kms$ in the southern lobe (blue) and in the range [$+40$;$+80$]$\kms$ in the northern lobe (red); c)~the southern terminal bow shock HH377, integrated in the range [$-77$,$-64$]$\kms$ (green); d)~the northern terminal bullet NB integrated in the velocity range [$+84$,$+97$]$\kms$ (magenta). First contour and contour interval are $20\%$ and $10\%$ of the peak intensity in each component, respectively. The synthesized beam  ($1\farcs07 \times 0\farcs87$, HPBW) is shown in the bottom left corner. The black dashed box draws the central protostellar region studied with the SO $N_J$= $5_4$--$4_3$ line (see Fig.~\ref{fig:CO-SO-North}). The main axis of the northern and southern lobes of the jet are shown with black arrows.} 
		\label{fig:CepE-CO}
	\end{center}
\end{figure} 

Cep\,E-mm  is an intermediate-mass Class 0 protostellar system of luminosity 100 L$_{\odot}$ \citep{Lefloch1996,Chini2001}, embedded in an envelope of $35\msol$ \citep{Crimier2010}, located in the Cepheus OB3 association at a distance of $819\pm 16\pc$ \citep{Karnath2019}. Using the NOEMA interferometer, \citet{Ospina-Zamudio2018} presented evidence of a protobinary system with a separation of $1\farcs35$ (1100~au) between component A located at $\alpha_{J2000}$= $23^h 03^m 12\fs8$, $\delta_{J2000}$= $+61\deg 42\arcmin 26\farcs0$ and component B at $\alpha_{J2000}$= $23^h03^m12\fs7$, $\delta_{J2000}$= $+61\deg 42\arcmin 24\farcs85$. While both protostars power a high-velocity jet,  the outflow from component A attracted attention because of its high luminosity in the near-infrared, in particular in the \htwo\ rovibrational line S(1) $\nu$= 1--0 at $2.12$~$\mu$m \citep{Eisloeffel1996}, a probe of shocked molecular gas. We will refer indistinctly to the driving  protostar as "A" or "Cep\,E-A" in what follows. 

High-velocity molecular jet emission was discovered for the first time by \citet{Lefloch1996} with the IRAM 30m telescope and subsequently imaged at $1\arcsec$ in the CO 2--1 line using the PdBI by \citet{Lefloch2015}. The CO 2--1 velocity-integrated emission of  the jet is displayed in Fig.~\ref{fig:CepE-CO}. The northern and southern lobes propagate at radial velocities of approximately $+65\kms$ and $-125\kms$, respectively, in the ambient cloud ($V_{lsr}$= $-10.9\kms$). Both lobes have similar length $\sim 20$--25$\arcsec$. The southern jet terminates with the bright Herbig-Haro object HH377 \citep{Ayala2000}, located at the tip of the molecular jet itself. In the north, the CO jet emission drops at $10\arcsec$ from the apex of the low-velocity outflow cavity (see Fig.~\ref{fig:CepE-CO}), where another, more extended "bullet", dubbed NB is detected near $V$= $+100\kms$. 

The proper motions of the \htwo\ bright shock structures, NB and HH377, were determined by \cite{Noriega-Crespo2014} and correspond to tangential velocities of $(70\pm 32)\kms$ and $(106\pm 29)\kms$ for the northern and southern components, respectively, after correcting for the recently revised distance to Cep\,E-mm (820 pc instead of 730 pc). Taking into account the radial jet velocity determination from the CO observations, this results in an revised jet inclination angle of $47^\circ$ with respect to the plane of the sky, close to the  determination by \citet{Lefloch2015}. 
The first indications of precession in the Cep\,E-mm outflow were reported by \citet{Eisloeffel1996}, who detected a wiggling structure and "sideways positional effects" in the \htwo\ 1--0 S(1) $2.12$~$\mu$m emission map of the outflow. The evidence for precession in the Cep\,E-mm outflow appears when inspecting the orientation of the two outflow cavities in the northern outflow lobe, which display a misalignement of $\approx 7\deg$, as can be seen in Fig.~\ref{fig:CepE-CO}. Interestingly, this value also corresponds to the difference between the current jet orientation (red in Fig.~\ref{fig:CepE-CO}) and the direction of the line connecting the northernmost knot NB ($\delta$= $+35\arcsec$) to protostar A.   

A simple model, assuming an outflow velocity of $200\kms$ and neglecting the inclination of the flow with respect to the plane of the sky, led \citet{Eisloeffel1996} to estimate an opening angle of  $4\deg$. More recently, based on numerical simulations, Noriega-Crespo \& Raga (2014) showed that the proper motions measured in the infrared in bowshocks could be explained by a jet precession angle of $10\deg$. Despite several parameters markedly different from more recent observational determinations (jet inclination angle of $30\deg$, assumed time-dependent velocity variations of 50\% over a $60\yr$ period), the authors qualitatively succeed in accounting for the overall features of the Cep\,E-mm outflow.

An accurate determination of the gas physical and dynamical properties (density, temperature, mass, momentum) in the low-velocity outflow and the jet were obtained  from a detailed CO multi-line analysis by \citet{Lefloch2015} and \citet{Ospina-Zamudio2019}. The authors showed that in the southern lobe the jet carries enough momentum ($1.7\msol\kms$) to accelerate the ambient gas and drive the low-velocity entrained gas ($2.6\msol\kms$). 
CO emission knots were found inside the southern jet, with sizes of $2\arcsec$--$4\arcsec$ and peak velocity variations of several $\kms$. They proposed that these knots trace internal shocks, which would be responsible for the hot (400--$750\K$) gas component detected inside the jet. 

As a conclusion, CepE\,-mm is a clear-cut case of a jet-driven protostellar outflow, whose in-depth study may shed new light on the formation and the dynamics of outflows/jets. 

\section{Observations}
\subsection{SO $N_J$= $5_4$--$4_3$}
The NOEMA array was used in configuration A with 8 antennas on 23 and 30 December 2016 to observe the spectral band 204.0--207.6 GHz towards the Cep\,E-mm protostellar system, as part of the Large Program SOLIS \citep[{\em Seeds Of Life In Space},][]{Ceccarelli2017}. The field of view of the interferometer was centered at position $\alpha$(J2000)= $23^{h}03^{m}13\fs00$ $\delta$(J2000)= $+61$\deg$42\arcmin21\farcs00$.
The wideband receiver WIDEX covered the spectral band 204.0--207.6 GHz  with a spectral resolution of 1.95 MHz ($\simeq 2.8\kms$), which includes the rotational transition SO $N_J$= $5_4$--$4_3$ at 206176.005 MHz. The spectroscopic parameters of the transition are taken from CDMS\footnote{CDMS database: https://cdms.astro.uni-koeln.de/cdms/} (M\"uller et al. 2005). 
The precipitable water vapor (PWV) varied between 1mm and 2 mm (between 2mm and 4mm in the second night) during the observations. Standard interferometric calibrations were performed during the observations: J2223+628 was used for both amplitude and phase calibration. Bandpass calibration was performed on 3C454.3. Calibration and imaging were performed by following the standard procedures with GILDAS-CLIC-MAPPING\footnote{http://www.iram.fr/IRAMFR/GILDAS/}. The emission of the SO $N_J$= $5_4$--$4_3$ line was imaged using natural weighting, with a synthesized beam of $0.52\arcsec \times 0.41\arcsec$ and a position angle PA of $359\deg$ (see Tab.~\ref{tab:solis_tech}). Self-calibration was applied to the data. 
The shortest and longest baselines were 72m and 760m, respectively, allowing us to recover emission at scales up to $12\arcsec$. The primary beam of the interferometer ($\approx 24.4\arcsec$) allowed us to recover emission up to $\sim 12\arcsec$ ($10^4$~au) away from protostars Cep\,E-A and Cep\,E-B, while the synthesized beam gave access to spatial scales of $\approx 400$~au. 

In order to estimate the fraction of the IRAM 30m flux collected by NOEMA, we compared the SO $N_J$= $5_4$--$4_3$ molecular spectrum of the ASAI millimeter line survey of the Cep\,E-mm protostellar core from \cite{Ospina-Zamudio2018} and the NOEMA spectrum degraded at the same angular resolution. We found that about 34\% of the IRAM 30m flux was recovered by the interferometer.  

\begin{table*}
\caption{Observational dataset properties}
%\begin{center}
\centering
\begin{tabular}{ccccccccc}%{lrrrrrccc}
\hline \hline
Line&Frequency&Backend&$\Delta\nu$&$\Delta V$& Config. &Synthetic beam&PA&rms\\
&(MHz)&(MHz)&(MHz)&($\kms$)&&($\arcsec \times \arcsec$)&($\deg$)&(mJy beam$^{-1}$)\\ \hline %\addlinespace
SO $N_J$= $5_4$--$4_3$& 206176.013 & WideX&2.0&2.8&A&$0.52\times 0.41$&$359$&2.2\\\hline %\addlinespace
CO 2--1&230538.000& WideX&2.5&3.3&AB&$1.07\times0.87$&$+73$&1.8\\\hline		
\end{tabular}\label{tab:solis_tech}
\tablefoot{The following properties are respectively presented: spectral bands, backend, spectral resolution $\Delta \nu$,  configuration of the interferometer, synthesized beam size, position angle PA, rms ($\mjy$~beam$^{-1}$) per element of spectral resolution $\Delta V$ ($\kms$).}
%\end{center}
\end{table*}

\subsection{CO 2--1}
In this work, we made use of the CO $J$=2--1 data  obtained at a spatial resolution of $\approx 1\arcsec$ with the IRAM Plateau de Bure Interferometer (PdBI). The observation procedure and the data were presented in \citet{Lefloch2015}. 
In order to map the molecular emission along the outflow, the 1.3mm band was observed over a mosaic of seven fields centered at the offset positions ($-9.0\arcsec$, $-20.8\arcsec$), ($-4.5\arcsec$,$-10.4\arcsec$), ($+0\arcsec$, $+0\arcsec$), ($+4.5\arcsec$,$+10.4\arcsec$), ($+4.5\arcsec$,$+20.8\arcsec$) with respect to Cep\,E-A.

As explained by the authors, the emission at scales larger than $10\arcsec$ was recovered from combining the PdBI data with the synthetic visibilities derived from an IRAM 30m map of the region. The jet total flux was recovered by the interferometer in both lobes \citep{Lefloch2015}. The intensities were converted from Jy/beam to K using the conversion factor of $26.88$.

\section{Jet-envelope interaction}

\begin{figure}
	\begin{center}

		\includegraphics[width=0.9\columnwidth]{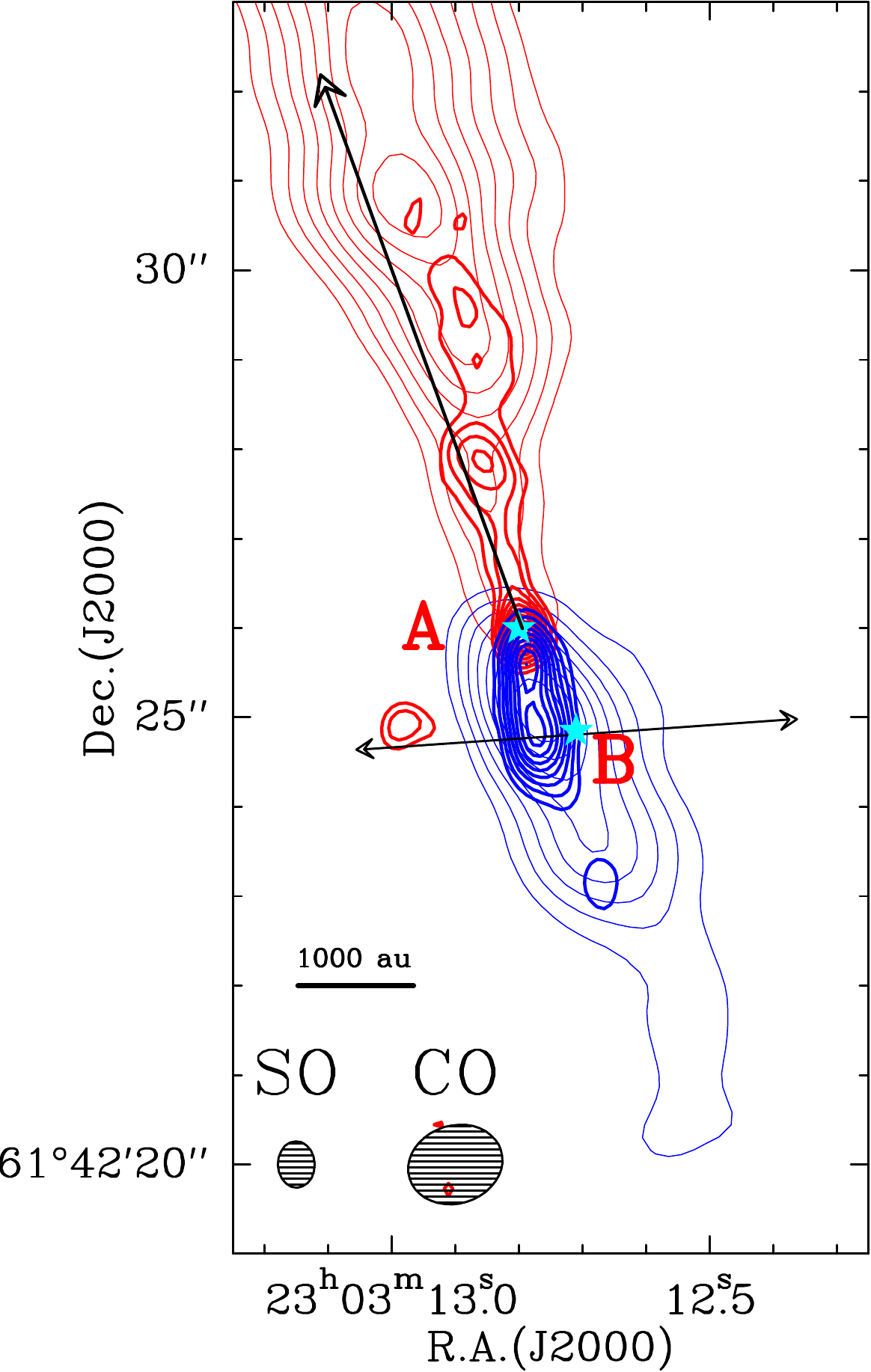} %normal 2 column
\caption{Distribution of the CO 2--1 and SO $N_J$= $5_4$--$4_3$ velocity-integrated emission in the central protostellar region. {\em Northern lobe (red):} the emission is integrated in the velocity range [+40;+80]$\kms$ and is displayed in thin and thick contours for CO and SO, respectively. {\em Southern lobe (blue):} the emission is integrated in the velocity range  $[-110;-50]\kms$ and is displayed in thin and thick contours for CO and SO, respectively. The velocity interval was chosen in order to include the region of jet acceleration close to the protostar (see Sect.~4.2). The beam size (HPBW) of the CO and SO observations ($1.07\arcsec \times 0.87\arcsec$ and $0.52\arcsec \times 0.41\arcsec$, respectively) are shown with ellipses in the bottom left corner. The velocity resolution of the CO (SO) observations is $3.3\kms$ ($2.8\kms$). First contour and contour interval are $20\%$ and $10\%$ of the peak intensity in each component. The arrow draws the main axis of the jet from protostar B \citep[from][]{Ospina-Zamudio2018}.} 
		\label{fig:CO-SO-North}
	\end{center}
\end{figure}

\begin{figure}
    \centering
       \includegraphics[width=0.9\columnwidth]{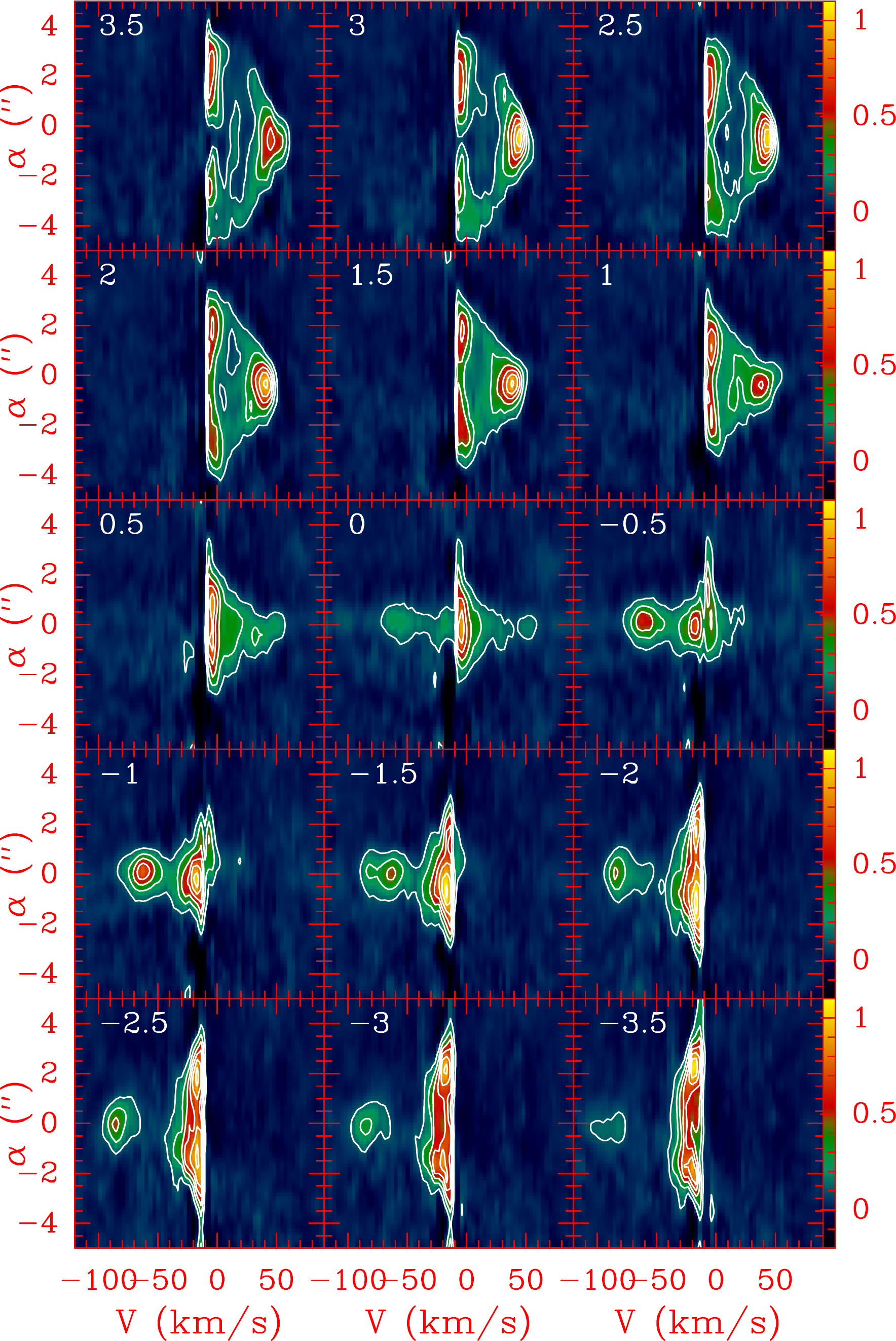} %2 column format
    \caption{Position-Velocity diagram of the CO 2--1 line emission  across the jet main axis (see Fig.~\ref{fig:CepE-CO}). The beam size (HPBW) of the CO observations is $1.07\arcsec \times 0.87\arcsec$. The velocity resolution of the  observations is  $3.3\kms$. The relative distance to the protostar $\delta$ (in arcsecond) is indicated in the top left corner. First contour and contour interval are 10\% and 15\% of the peak flux, respectively. Fluxes are expressed in Jy/beam.}
    \label{fig:PV_CO_across}
\end{figure}

\subsection{Molecular gas distribution}
The CO and SO emissions from the high-velocity jet in the protostellar region around Cep\,E-A are displayed in Fig.~\ref{fig:CO-SO-North}. The line fluxes were integrated in the velocity range [$+40$;$+80$]$\kms$ and [$-110$;$-50$]$\kms$ in the northern and southern lobe, respectively. 
We observe a very good morphological match between the SO and CO emission distributions at close distance  ($< 2\arcsec$) from the protostar. Both species trace a strongly collimated jet with a diameter $\leq 0.4\arcsec$ ($\sim 320$~au), unresolved in our observations. At a larger distance ($ > 2\arcsec$) from protostar A, the CO emission gradually broadens until it reaches a diameter of about $2\arcsec$ (FWHM) while the SO emission is mainly unaffected. Comparison of CO Position-Velocity diagrams across the jet main axis in Fig.~\ref{fig:PV_CO_across} confirms that the high-velocity emission ($V$= $60\kms$) is unresolved near the protostar (offset $\delta= 0\arcsec$) while it reaches a diameter  of $2\arcsec$ at $\delta > 1.5\arcsec$.

Hence, these observations draw the picture of a two component jet: a central, highly-collimated, component of diameter $< 320$~au, specifically probed by SO, surrounded by a radially extended gas layer detected in CO. The pattern revealed in our observations is consistent with the one recently reported by \citet{Podio2021} in the CALYPSO sample of young low-mass protostellar outflows.

The situation is more complicated in the southern lobe because of the high-velocity jet powered by protostar B, which propagates in the eastern-western direction at $1\arcsec$ South of protostar A \citep{Ospina-Zamudio2018}. A signature of the B jet is detected as a redshifted clump near $23^h 03^m 13\fs0$ $+61\deg 42\arcmin 25\arcsec$ (Fig.~\ref{fig:CO-SO-North}). We note that the southern (blueshifted) lobe of the A jet displays a highly-collimated SO emission, with a diameter similar to that of the northern lobe.

\subsection{Jet acceleration}
The orientation of the main axis of the jet at large scale was determined from the distribution of the high-velocity CO emission integrated in the intervals $[+40;+80]\kms$ and $[-135;-110]\kms$ in the northern and southern lobe, respectively (see Fig.~\ref{fig:CepE-CO}).  The former is found to make a position angle (PA) of $+20\deg$ while the latter makes a PA of $+204\deg$. These values based on the large-scale CO distribution are somewhat indicative because of the complexity of the outflow kinematics and the evidence of changes in the direction of propagation along the outflow, as illustrated at large scale by the difference of orientation of the two northern outflow cavities (Fig.~\ref{fig:CepE-CO}). This point will be discussed in Sect.~6. 

\begin{figure}[!ht]
\centering
\includegraphics[width=0.93\columnwidth]{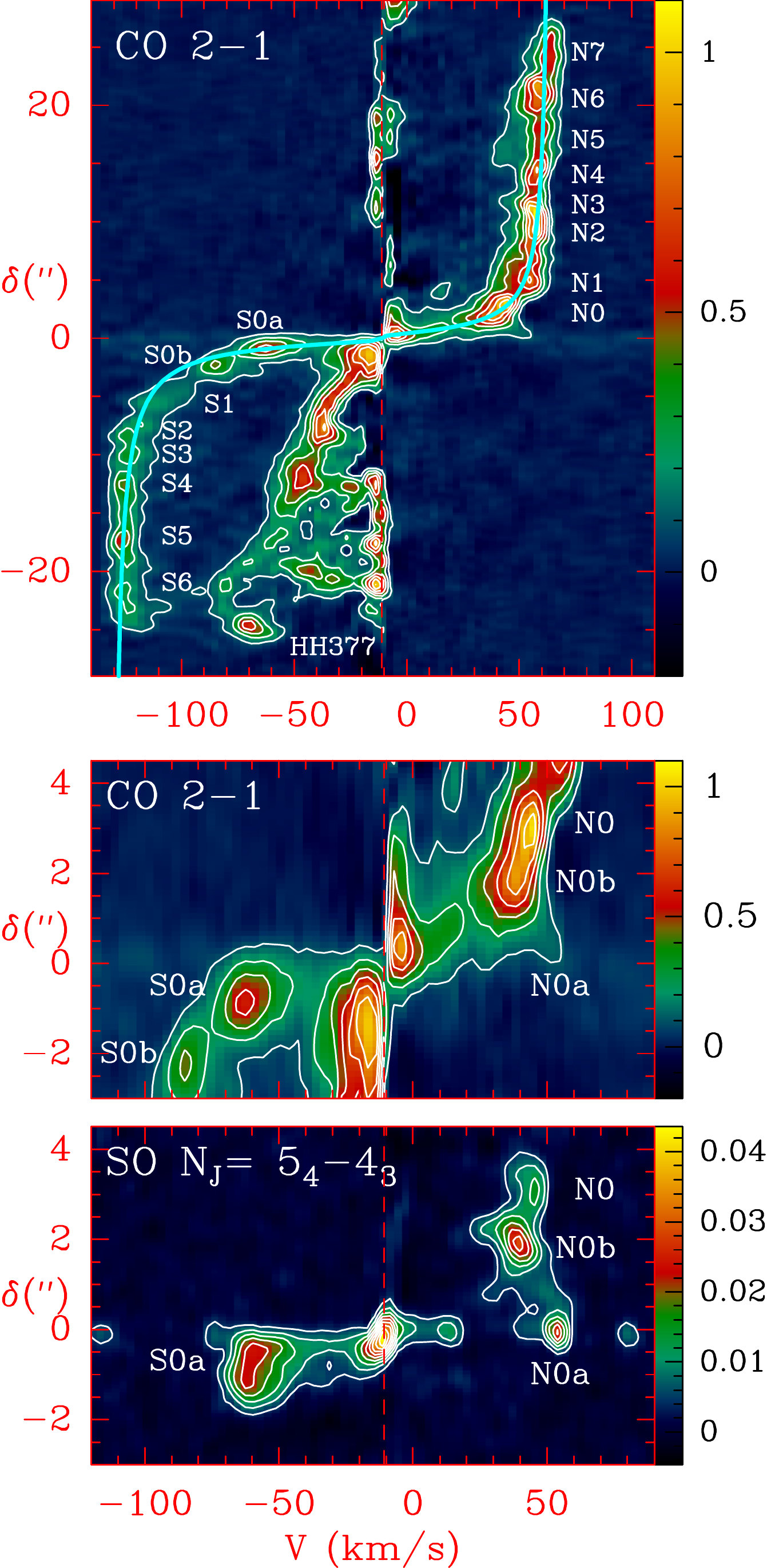} 
\caption{Position-velocity diagrams of the CO and SO emission along the jet main axis (see Fig.~\ref{fig:CepE-CO}). {\em (top)}~CO 2--1 as observed at $1\arcsec$ with the IRAM interferometer. The different knots identified close to the protostar are indicated. The best fitting relation to the gas acceleration |$V-V_{lsr}$|/$V_0$= $\exp(-\delta_0/\delta)$ is superimposed in cyan. {\em (Middle)}~Magnified view of the CO 2--1 emission in the central protostellar region near Cep\,E-A. 
{\em (bottom)}~SO $N_J$= $5_4$--$4_3$  as observed at $0.4\arcsec$ resolution with NOEMA. First contour and contour interval are 10\% and 15\% of the peak intensity, respectively. The red dashed lines mark the ambient cloud velocity ($V_{lsr}= -10.9\kms$).
The velocity resolution of the CO and SO observations is  $3.3\kms$ and $2.8\kms$, respectively. Fluxes are expressed in Jy/beam.}
    \label{fig:PV_CO_SO_Main}
\end{figure}

The distribution of the CO emission as a function of velocity along the main jet axis is displayed in Fig.~\ref{fig:PV_CO_SO_Main}. The emission is dominated by the protostellar source Cep\,E-A at $V_{lsr}$= $-10.9\kms$. At large distance from the source ($|\delta| > 10\arcsec$), the jet radial velocity $V$ reaches a steady value of $\approx +65\kms$ and $-125\kms$ in the northern and southern lobe, respectively. 

However, in the inner $5\arcsec$ ($\approx 4000$~au) around the protostar, the CO jet displays velocities much lower than the asymptotic values measured at large distances. More precisely, the velocity gradually increases from ambient at $V_{lsr}=-10.9\kms$ to the terminal value on a scale of a few arcseconds. The radial velocity profile $V$ as a function of distance $\delta$ to the protostar along the jet main axis can be fitted by the simple relation: 
\begin{equation*} 
   (V-V_{lsr})/V_0 = \exp(-\delta_0/\delta),
\end{equation*}
where $V_0$ is the jet asymptotic radial velocity {\em relative to the protostar} and $\delta_0$ the length scale of the jet acceleration process.  %\noident here

The length scale  $\delta_0$ and the jet velocity $V_0$ (relative to Cep\,E-A) were determined from a best fitting procedure to the CO Position-Velocity diagram. The procedure yields $\delta_0$= 693~au (590~au) and a jet velocity $\rm V_0$= $+75\kms$ ($-118\kms$) in the northern (southern) jet. The agreement between the best fitting $V_0$ values and the  observational determination of the jet velocity relative to the source is very good ($+76\kms$ and $-114\kms$ for the northern and southern lobes, respectively; see Sect.~2 and Fig.~4) and supports our simple modeling. The close agreement between both values of $\delta_0$ suggests that the physical conditions are  similar at subarcsec scale around Cep\,E-A.  

Intriguingly, the jet appears to accelerate on a scale of 700 au, exceeding by far  the size of the protostellar disk and hot corino of Cep\,E-A $\approx 200$~au (Lefloch et al. {\em in prep}). Inspection of the CO and SO Position-Velocity diagrams along the major axis of the jet in the central protostellar regions reveal a more complex situation, as can be seen in the middle and bottom panels of Fig.~\ref{fig:PV_CO_SO_Main}. The CO emission distribution reveals that a small fraction of material is actually accelerated at velocities up to $\approx 60\kms$ at a few 100~au from protostar A, hence similar to the jet terminal velocity. 

The SO $N_J$= $5_{4}-4_{3}$ emission distribution obtained at $0.4\arcsec$ resolution, a factor of 2 higher than in the CO observations, brings more insight into the gas acceleration as can be seen in Fig.~\ref{fig:PV_CO_SO_Main} (bottom panel).
The bulk of the SO emission arises from three compact knots: N0a, located at less than $0.2\arcsec$ ($\sim 160$~au) from Cep\,E-A, with velocity $V$= $+54.6\kms$, and N0b and N0, detected in  the northern jet at $\delta$= $+2\arcsec$, $V$= $+39.8\kms$ and $\delta$= $+3.1\arcsec$, $V$= $+44.5\kms$ respectively.  As can be seen in Fig.~\ref{fig:CO-SO-North},  the SO emission in the velocity range of these three knots ($[+40;+80]\kms$) is unambiguously located in the northern lobe of the Cep\,E-A jet. 
As shown by \citet{Ospina-Zamudio2018}, the  second high-velocity, collimated jet powered by the secondary companion Cep\,E-B propagates in the eastern-western direction about $1\arcsec$ South of Cep\,E-A (see their Fig.~2), nearby enough to rise the question of a possible contamination in the molecular gas emission distribution. 

However, as can be seen from the spatial distribution of the SO emission at $0.4\arcsec$ resolution integrated between $+40$ and $+80\kms$ (Fig.~\ref{fig:CO-SO-North}), all three knots are associated with the Cep\,E-A northern jet. Interestingly, the bulk velocity of the SO gas appears to vary monotonically between these knots. 

To summarize, the analysis of the jet kinematics reveals the presence of two components: an extremely high velocity, highly collimated component originating from the protostar, detected in SO and CO, and a secondary, more extended layer of accelerated gas detected mainly in CO. 

\subsection{Envelope entrainment}

\subsubsection{Northern lobe}
Inspection of the CO Position-Velocity diagrams across the jet main axis (Fig.~\ref{fig:PV_CO_across}) shows that the ambient gas emission extends up to distances of $\pm 4\arcsec$ from the jet (see e.g. panel $\delta$= $+2\arcsec$). 
The CO jet is well identified at  $V\simeq$ $+65\kms$. In the close environment of the protostar at $\delta= 0\arcsec$, one observes a continuous velocity distribution between the outer quiescent regions of the envelope and the jet, as can be seen in Figs.~\ref{fig:PV_CO_across}--\ref{fig:PV_CO_SO_Main}. This means that the ambient, quiescent protostellar material from the envelope is dragged away and accelerated by the jet. 
The CO emission of this entrained component is brighter than the jet material launched from the protostar and its associated knots, hence suggesting  an accelerated jet. The gas entrainment is best seen in the PV diagrams  $\delta= +0\arcsec$ to $\delta$= $+1.5\arcsec$  across the jet main axis (see top panels in Fig.~\ref{fig:PV_CO_across}). At larger distances from the protostar, $\delta \geq 2.0\arcsec$, the jet signature ($V$= $+65\kms$) gradually separates from the entrained gas emission. This suggests that the entrainment process loses its efficiency as the distance from the protostar increases beyond  $\sim 2\arcsec$ (1600~au).  

\subsubsection{Southern lobe} 
The evidence of jet entrained gas near Cep\,E-A at $\delta$= $0.0$ is much weaker. The emission in the range [$-70$;$-11$]$\kms$ appears collimated with a typical diameter of $1.5\arcsec$, consistent with the value measured in the northern jet in the range [$+40$;$+60$]$\kms$. The main difference with the northern lobe lies in that there is no jet entrainment of material at distance $> 1\arcsec$ (800~au), i.e. that the amount of jet entrained material is much less in the southern lobe.
The CO jet emissivity is inhomogeneous and displays a maximum between $\delta$= $-1\arcsec$ and $\delta$ = $-3\arcsec$ (see Fig.~\ref{fig:PV_CO_across}). This maximum coincides with the region of propagation of the Cep\,E-B jet, as mapped in the SiO 5--4 line by \citet{Ospina-Zamudio2019}. We speculate that the absence of gas entrainment in the southern lobe could result from the interaction between both outflows/jets. 

As discussed by \cite{Lefloch2015}, detailed analysis of the momentum budget in the outflow southern lobe showed that the high-velocity jet carries away the same amount of momentum as the low-velocity gas (1.7 vs $2.6\msol\kms$, respectively). This led the authors to conclude that the low-velocity outflow is driven by the protostellar jet itself. This result is supported by the distribution of the CO emission in the Position-Velocity diagram along the main jet axis. As can be seen in the top panel of Fig.~\ref{fig:PV_CO_SO_Main}, the low-velocity gas appears to accelerate gradually from the protostar up to the terminal shock HH377 following a "Hubble law" relation: $(V-V_{lsr}) \propto \delta $ where $\delta$ is the angular distance to the protostar. This kinematical feature is an unambiguous signature of jet bowshock driven outflow \citep{Raga-Cabrit1993,Wilkin1996,Cabrit1997}.

\section{Mass-ejection}

The presence of bright knots of CO gas in the molecular jet was first reported by \citet{Lefloch2015}. In this section, we present a detailed study of the knots and we show how a simple ballistic modeling can account for their dynamical properties.

\subsection{Kinematics}\label{knot-properties}

We have searched for knots along the jet in both the northern and the southern lobes from the CO jet intensity map in Fig.~\ref{fig:CepE-CO} and the CO Position-Velocity diagram along the jet main axis (Fig.~\ref{fig:PV_CO_SO_Main}). We identified ten (eight) CO knots in the northern (southern) lobe,  labeled N0a to N7 (S0a to S6), respectively (see Table~\ref{tab:knots}).

The SO $5_4$--$4_3$ observations provide a more concentrated view of the knots located at a few arcsec from the protostar.  In the northern jet, we have detected knot N0a at $\delta$= $+0.2\arcsec$ and $V$= $+54.6\kms$ and knot N0b at $\delta$= $+2.0\arcsec$ and $V$= $+39.8\kms$ (see bottom panel in Fig.~\ref{fig:PV_CO_SO_Main}).
In the southern jet, we have detected a knot, referred to as S0a,  near offset position $\delta$= $-0.8\arcsec$ (and $V$= $-62.5\kms$). We note that the SO emission looks elongated along the jet axis and it is hence unclear whether more than one knot is actually present.
The emission from the southern jet close to Cep\,E-A does not display the same fragmented structure as in the north.  The similarity in the ejection velocities of  S0a and N0a, both at short distance from Cep\,E-A suggests that they are associated with the same ejection event and that the physical conditions of jet acceleration are comparable in both lobes.

We have derived the main physical properties of the identified knots: coordinates, size (beam-deconvolved), distance $d$ from the protostar, radial velocity $V$, dynamical age $t_{dyn}$= $d/V \times \tan i$, and mass.  
The positions and sizes of the different knots were determined from a 2D gauss fit to the CO flux distribution, except for knots N0a-N0b, whose parameters were determined from SO. The velocities were measured on the CO Position-Velocity diagram in Fig.\ref{fig:PV_CO_SO_Main}. 
Those parameters are summarized in Table~\ref{tab:knots}. For the sake of completeness, we have included the parameters of the terminal shocks NB and HH377 (see Fig.~\ref{fig:CepE-CO}). 

The knot dynamical ages  appear to span the time range 60--$1500\yr$. Interestingly, HH377 and NB have comparable dynamical ages, so that both series of knots in the northern and southern appear to cover the same time interval of mass-ejection events. 

\begin{figure}[!ht]
\centering
\includegraphics[width=0.48\textwidth]{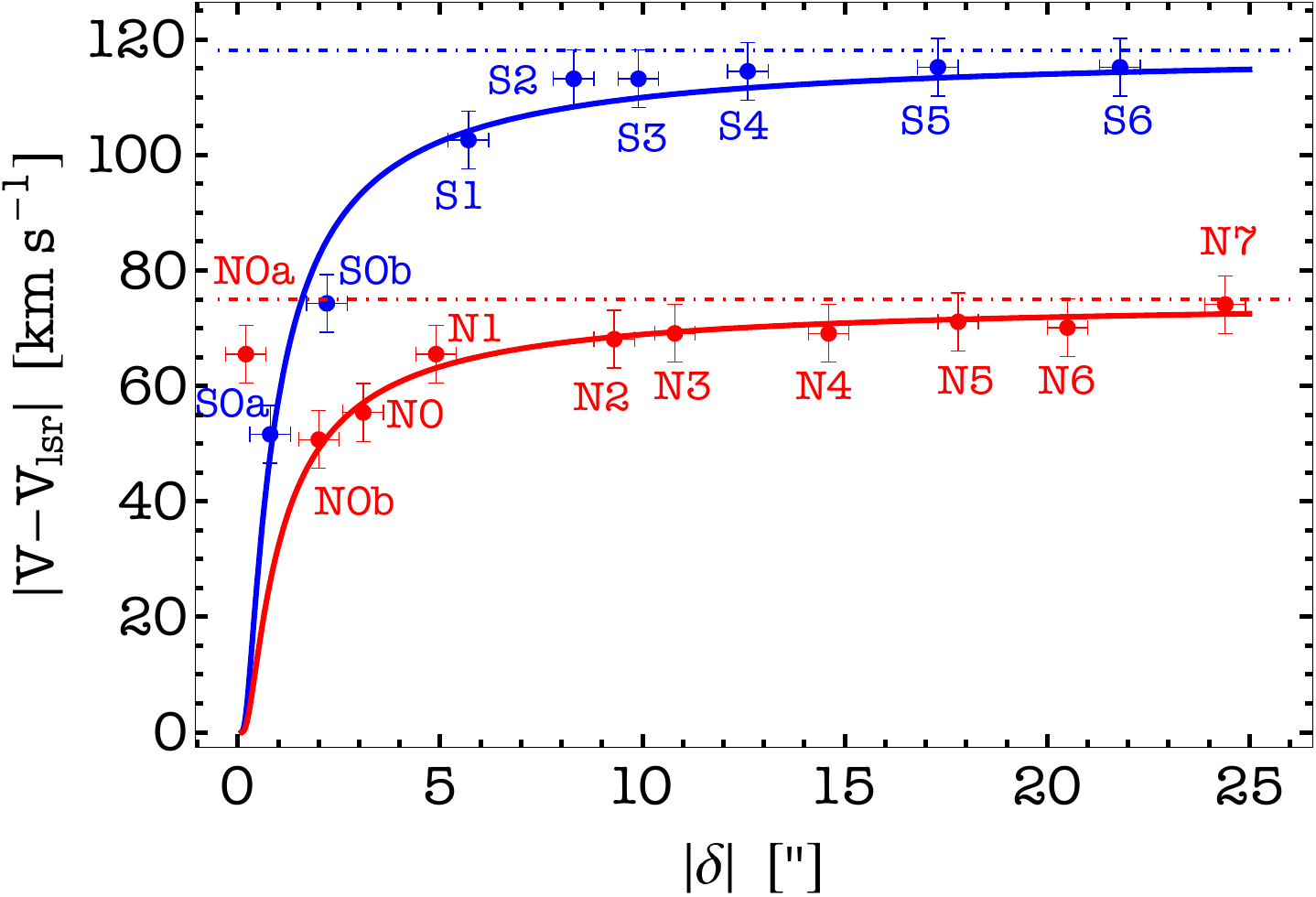}
\caption{Distribution of knot velocity peaks along the main jet axis as a function of distance $\delta$ to Cep\,E-A in the northern (red) and southern (blue) jet components. The best fit to the northern (southern) jet velocity relative to the source  $(V-V_{lsr})/V_0$= $\exp(-\delta_0/\delta)$  with $\delta_0= 693$~au ($\delta_0$= 590~au) and $V_0$= $+75\kms$ ($V_0=-118\kms$) is drawn by the solid curve.  }
    \label{fig:PV-fragments}
\end{figure}

Figure~\ref{fig:PV-fragments} displays the distribution of knot velocities relative to Cep\,E-A as a function of the distance to the protostar. We have superimposed the best fitting solutions to the CO Position-Velocity diagrams in blue and red  for the southern and northern lobes, respectively. We adopted an uncertainty of $\pm 0.5\arcsec$ in position (one beam size) and a  value of $\pm 5\kms$ in velocity (mean knot linewidth) (see Table~\ref{tab:knots}). It immediately follows that all the knots but N0a trace the overall CO gas distribution in the jet. The amplitude of the velocity fluctuations with respect to the bulk of jet material velocity is weak, of the order of a few $\kms$.  

\begin{figure}[!h]
\centering
\includegraphics[width=0.48\textwidth]{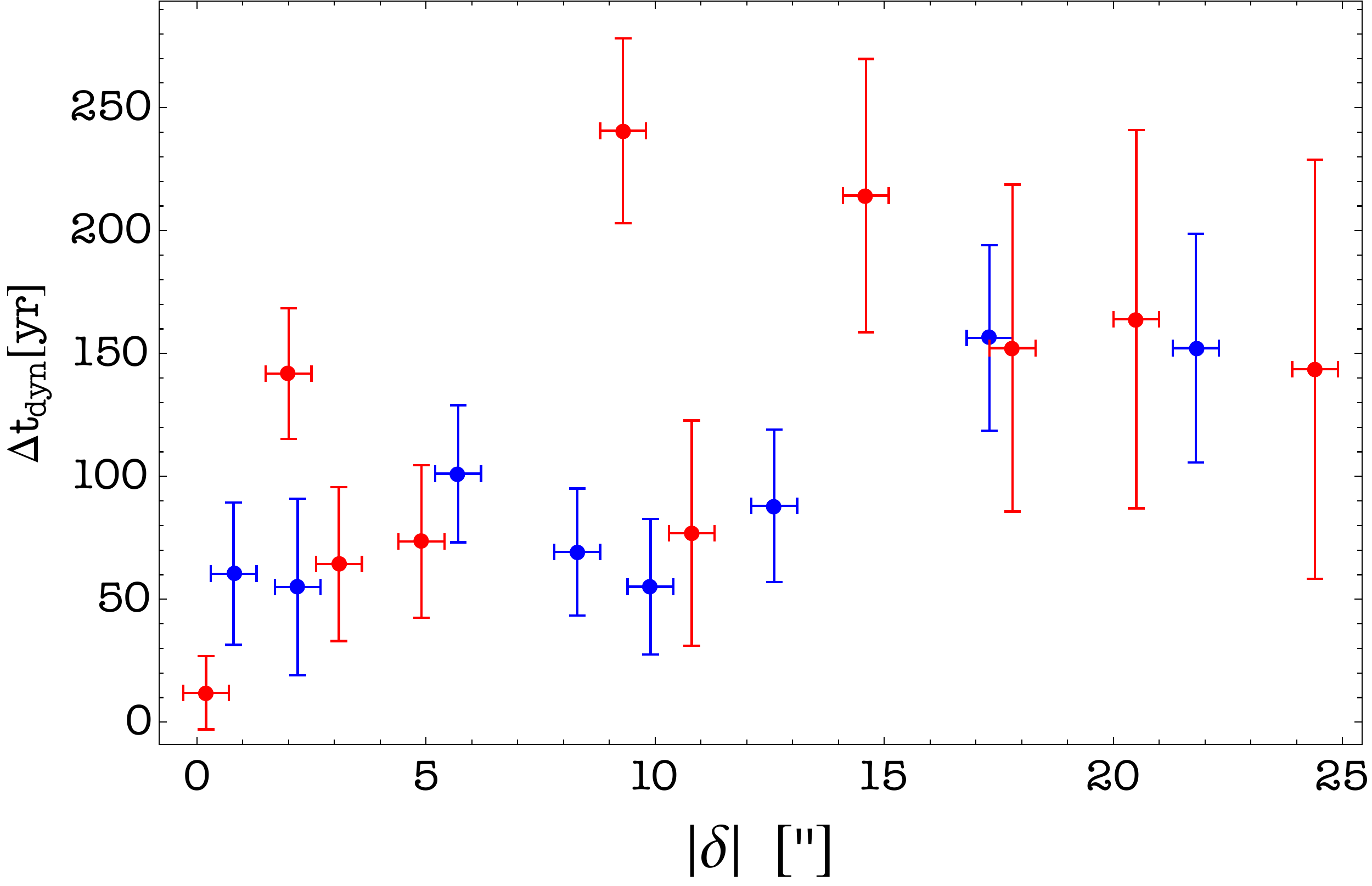}
\caption{Distribution of time intervals between subsequent knots as a function of distance to the protostar along the jet main axis, both in the northern (red) and southern (blue) lobes.}
    \label{fig:time-interval-d}
\end{figure}

Knots are not regularly distributed along the jet, at equal distance one from the other. At large distance from the protostar, typically $10\arcsec$, the separation between knots tends to increase from $\sim 2.5\arcsec$ to $\sim 4.5\arcsec$ in the southern lobe (see Fig.~\ref{fig:PV-fragments}). The knot distribution is more complex in the northern lobe,  but their separations are of the same order, $\sim 2.5\arcsec$ (2000~au) and $4.5\arcsec$ (3700~au) for "close" (e.g. N2/N3) and "remote" (N1/N2) knots, respectively. This distribution may be biased by the gas acceleration in the central protostellar regions.  For this reason, in order to get more insight into the knot distribution, we have computed the distribution of time intervals between subsequent knots as a function of distance to the protostar. It is displayed in Fig.~\ref{fig:time-interval-d}. A marked separation is observed in the southern lobe (in blue) between the knots located at  $\leq 10-13\arcsec$ from the protostar, with a short $\Delta t_{dyn}$ $\sim 50$--$80\yr$ , and the more distant knots, with $\Delta t_{dyn}$ $\sim 150$--$200\yr$. A similar pattern is observed in the northern lobe (red), with similar time intervals close to and far away from the protostar.  We consider the case of knots N2/N3 at $9$--$11\arcsec$ as peculiar; we suspect this anomaly is caused by the interaction of the knots with the  ambient gas.

To summarize, analysis of the knot interval distribution suggests the presence of two timescales: a short time interval $\Delta t \sim 50-80\yr$ close ($\leq 12\arcsec$) to the protostar and a longer timescale $\Delta t \sim 150-200\yr$ at larger distances from the protostar. Such knot interval distribution could be explained due to a periodic process taking place at the base of the Cep\,E-mm jet. We examine this hypothesis in more detail in the next section. 

\subsection{Knot mass distribution}

\begin{figure}[!h]
\centering
\includegraphics[width=0.48\textwidth]{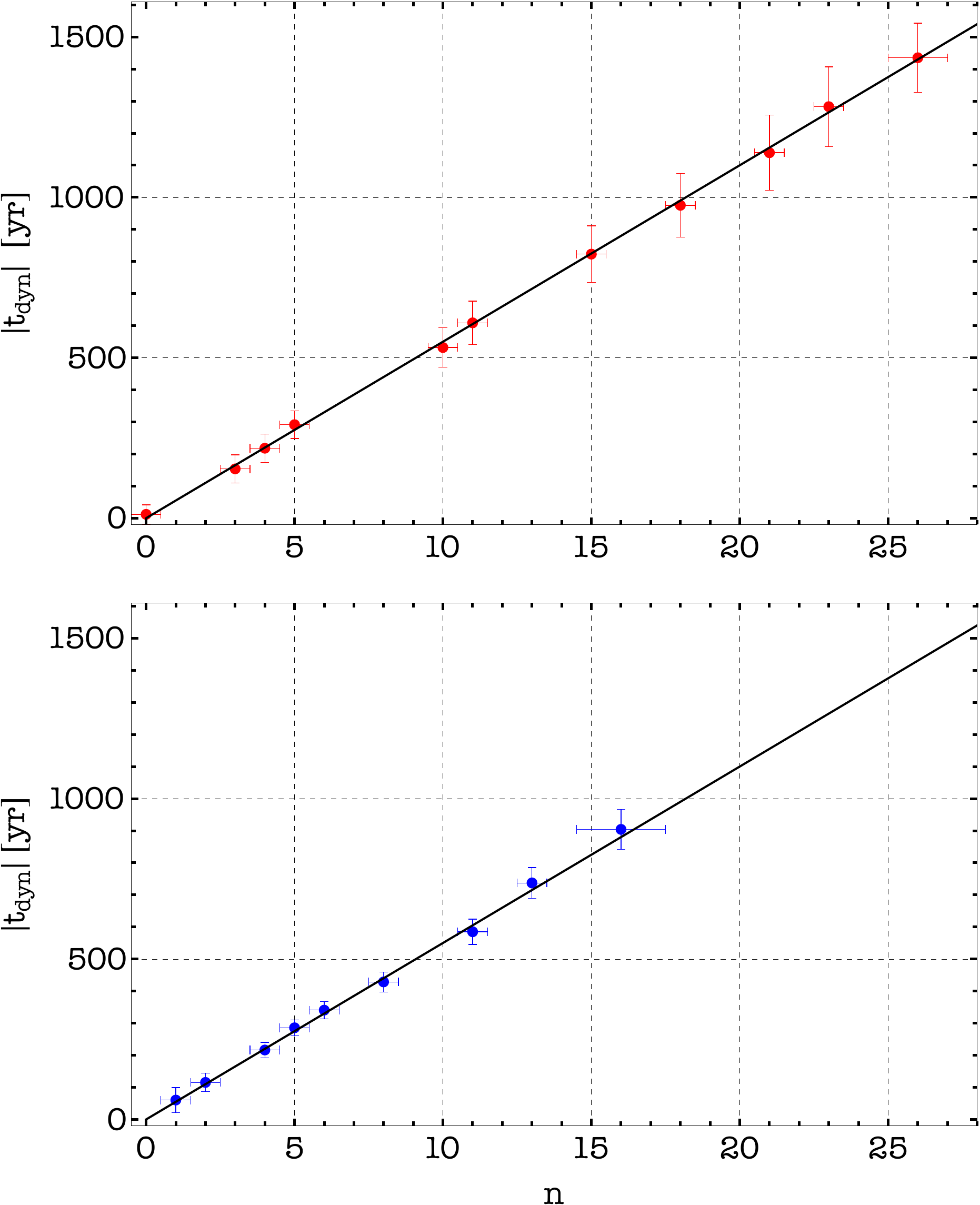}
\caption{Distribution of knot dynamical ages as a function of the number of periods of the periodic mass-ejection mechanism $n$, in the northern (upper panel) and southern (lower panel) lobes, respectively.}
    \label{fig:age-knots}
\end{figure}

We have determined the mean gas column density of the knots from a radiative transfer analysis of the CO 2--1 line emission, using the code MadeX \citep{Cernicharo2012} in the Large Velocity Gradient approximation and the CO-\htwo\ collisional coefficients from \cite{Yang2010}. In this simple model, we assume that the CO emission measured towards the knots actually arises from a single physical component. In other words, the contributions of the entrained gas envelope and the inner jet material along the line of sight, both discussed in Sect.~4.3, are neglected with respect to that of the knots.  Linewidths and intensities were obtained from a Gaussian fit to the line profiles at the knot flux peak (see Table~\ref{tab:knots}). The knot peak intensity was sometimes found to lie off the jet main axis by a fraction of a beam. This indirectly reveals a complex knot structure at subarcsec scale, with some small velocity shifts between the intensity peak and the jet main axis.  
We have adopted the jet \htwo\ density and temperature derived by \cite{Lefloch2015} and \cite{Ospina-Zamudio2019} in the southern and northern lobes from a CO multi-transition analysis.
The total \htwo\ mass and column density were subsequently obtained from N(CO) adopting a standard CO to \htwo\ relative abundance ratio of $10^{-4}$. 

We derive knot masses of the order of $10^{-3}\msol$ (see Table~\ref{tab:knots}). The knots are somewhat more massive in the north than in the south by a factor of a few. In both lobes, the masses of the knots tend to increase beyond $10\arcsec$ (8000~au), a distance corresponding to knots N3 and S4 in the northern and southern lobe, respectively, or, equivalently, to a dynamical timescale of $500\yr$. As discussed in Sect.~4.3, the CO Position-Velocity diagrams show evidence of  protostellar material entrainment by the northern jet only in the central $2\arcsec$--$3\arcsec$ close to Cep\,E-A; therefore, it seems unlikely that the observed knot mass increase is related to the jet-envelope interaction. This point is addressed in more detail in the following section, in which we consider the possibility of knot dynamical interactions. 

The mass of  knot NB ($1.3\times 10^{-3}\msol$) is similar to that of the other knots, in agreement with the interpretation of NB being the jet terminal bowshock. Our estimate of the HH377 mass was found in good agreement with the previous estimate obtained by \cite{Lefloch2015} ($\simeq 6\times 10^{-3}\msol$), based on a detailed CO multi-line analysis. In other words, the mass of HH377 amounts to that of 4--5 knots, consistent with  the number of "undetected" knots in the southern lobe. The derived value must be considered as a lower limit as the CO abundance could be lower if dissociative J-type shock components are present \citep[see][]{Gusdorf2017}.

\begin{table*}[th]
\begin{center}
{\small
\hfill{}
\caption{Knot identification and properties}
\begin{tabular}{ccccccccccc} \hline \hline
ID &($\Delta\alpha$;$\Delta\delta$)& $\delta$ & Size & $V$&$\Delta V$& F$_P$ & $t_{dyn}$&$n$ & Mass\\
   &($\arcsec$;$\arcsec$)&($\arcsec$)&($\arcsec$) & ($\kms$)&($\kms$)&(K $\kms$) &(yr)& & ($10^{-3}\msol$)\\
   \hline%\addlinespace
North   &     &      &      & \\
N0a&($+0.1;+0.2$)& 0.2  & 0.5 & +54.6 & 12(3) & 35(9) &12 & 0 &0.08\\
N0b&($+0.3;+2.0$)& 2.0  & 0.9 & +39.8 & 18.3(0.4) & 430(9) &154 &  3 &0.3\\
N0 &($+0.6;+3.1$)& 3.2  & 1.4 & +44.5 & 14.1(0.3) & 289(7) &218 &  4 &1.9\\  
N1 &($+1.2;+4.7$)& 4.9  & 1.8 & +54.6 & 18.7(0.7) & 351(13) &291 &  5 &3.6\\
N2 &($+2.3;+9.0$)& 9.3  & 1.7 & +57.2 & 9.8(0.4) & 270(10) &532 & 10 &2.8\\
N3 &($+3.0;+10.4$)& 10.8 & 3.0 & +58.2 & 8.5(0.4) & 261(12)&609 & 11 &7.8\\
N4 &($+4.6;+13.9$)& 14.6 & 2.6 & +58.2 & 9.6(0.4) & 197(7)&823 & 15 &4.5\\
N5 &($+5.4;+16.9$)& 17.8 & 2.0 & +60.2 & 12.5(0.9) & 189(14)&975 & 18 &2.5\\
N6 &($+6.6;+19.4$)& 20.5 & 2.9 & +59.2 & 13.3(0.3) & 281(6)&1139& 21 &7.6\\
N7 &($+8.3;+22.9$)& 24.4 & 1.9 & +63.2 & 10.2(0.3) & 175(5)&1283& 23 &2.1\\ \hline
NB &($+6.3;+34.3$)& 34.9 & 4.0 & +83.3 &13(2) & 70(9) &1440& 26 &1.3 \\
%   &      &     &       & 1660$\dag$& \\
%N8 & $+7.7$ & $+40.0$ & 40.7 & 4 \\
\hline%\addlinespace
South  &         &       &      &  & \\
S0a &($-0.2;-0.8$)& -0.8  & 0.9 & -62.5  & 23.6(0.5)& 389(9)&60 & 1 & 0.7\\
S0b &($-0.8;-2.1$)& -2.2  & 1.1 & -85.2  & 19(1)& 216(16)&115& 2 & 0.6\\
S1 &($-1.8;-5.4$)& -5.7  & 1.3 & -113.5  & 20(2) & 147(11)&218& 4 &0.5\\
S2 &($-3.3;-7.6$)& -8.3  & 1.7 & -124.1  & 16.0(0.8) & 135(7)&286& 5 &0.8\\
S3 &($-3.6;-9.3$)& -9.9  & 1.8 & -124.1  & 14.7(0.7) & 144(7)&341& 6 &0.9\\
S4 &($-4.9;-11.6$)& -12.6 & 2.3 & -125.4  & 8.4(0.3) & 114(4)&429& 8 &1.3\\
S5 &($-6.8;-15.9$)& -17.3 & 2.0 & -126.1  & 8.4(0.3) & 124(5)&585& 11 &1.1\\
S6 &($-8.8;-19.9$)& -21.8 & 1.4 & -126.1  & 13.2(0.3) & 114(6)&737& 13 &0.5\\ \hline
HH377 &($-9.5;-22.7$)& -24.6 & 2.2& -67.0 & 10.9(0.2) & 261(5)&904& 16 & 6.0\\	\hline  
\label{tab:knots}
\end{tabular}
}
\hfill{} \\	
\tablefoot{Coordinates, distance to Cep\,E-A along the jet main axis, size (beam deconvolved), radial velocity, line width, flux at the peak, dynamical age, period number and mass of the knots in the jet. The dynamical ages for NB and HH377 were directly computed from the proper motions measured by \citet{Noriega-Crespo2014} after correcting for the difference of inclination angle induced by the jet precession.}
\end{center}

\end{table*}

\subsection{ Mass-loss rate}

\begin{figure}[!ht]
\centering

\includegraphics[width=\columnwidth]{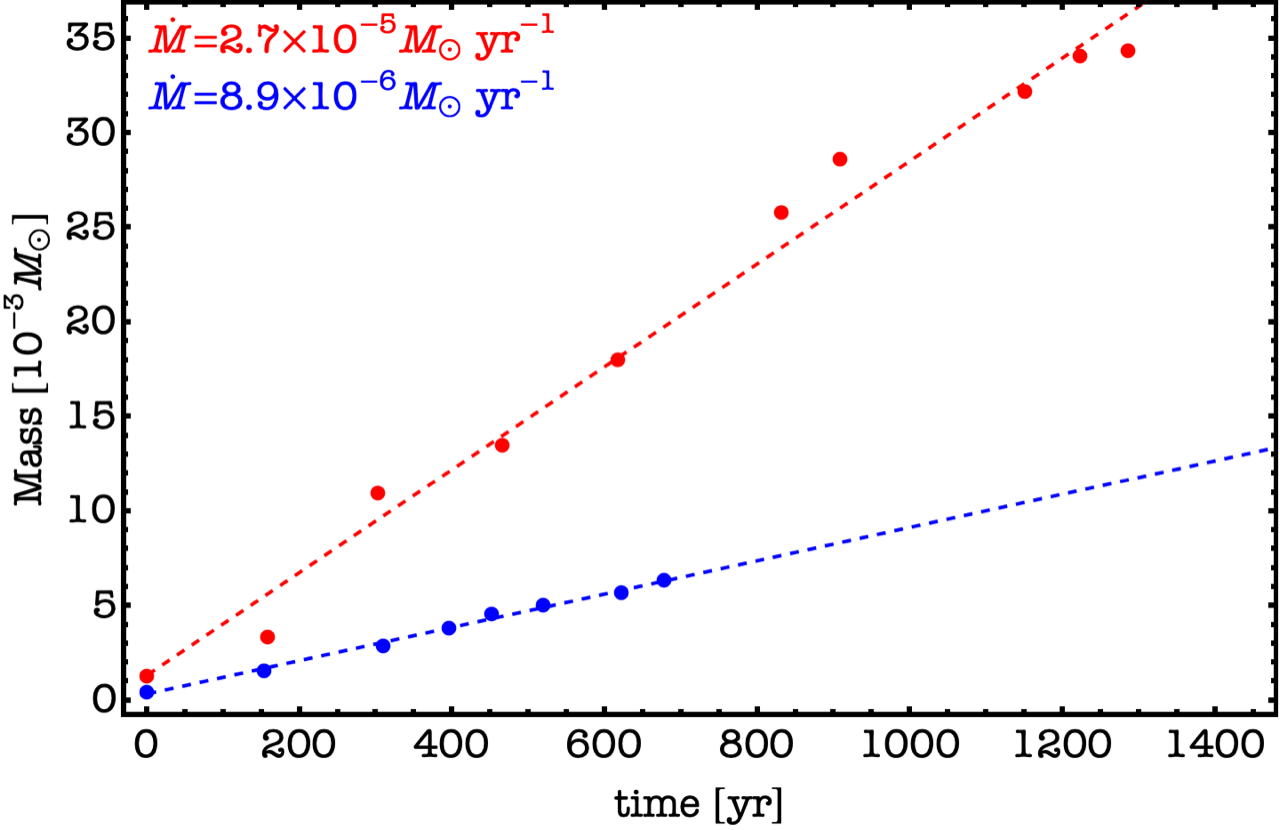}
\caption{Cumulative knot mass function in the northern (red) and southern (blue) lobes. Knot NB (S6) and its associated dynamical timescale is taken as reference in the northern (southern) lobe. The best linear fits are  drawn by dashed lines.}
    \label{fig:mass-loss-rate}
\end{figure}

The determination of knot masses and their dynamical timescales allows us to retrieve the mass-loss history of Cep\,E-A and its variations. We report in Fig.~\ref{fig:mass-loss-rate} the cumulative mass of ejected knots as a function of time. In this analysis, the origin of the time axis {\em t=0} is defined by the knots NB and S6 in the northern and southern lobes, respectively (strictly speaking, HH377 is not a simple knot, as discussed in the next Section). Then, the cumulative mass is obtained by adding up one after the other the mass of the subsequently ejected  knots; the corresponding time is simply obtained from the distribution of knot dynamical timescales, starting again from the most distant ones (NB/S6). We note that the reference time "t=0" in Fig.~\ref{fig:mass-loss-rate} differs between outflow lobes as the reference knots NB and S6 taken as reference to compute the cumulative mass function  have different dynamical ages. 

As can be seen in Fig.~\ref{fig:mass-loss-rate}, the cumulative mass function is well fitted by a linear function of slope $\dot{M}$= $8.9\times 10^{-6}\msol\yrmu$ in the southern lobe (blue dashed line) and  $\dot{M}$= $2.7\times 10^{-5}\msol\yrmu$ in the northern lobe (red dashed line), respectively.  We can draw two conclusions from this figure. 

First, the mass-loss rate appears steady  over the whole knot ejection process in both lobes. 
We note that the possibility of knot interactions along the jet, as discussed in Sect.~5.4, should not affect the slope of the cumulative mass function but only the mass of specific ejecta.  

Second, the mass-loss rate appears to be higher in the northern lobe by about a factor 3. The main difference between both lobes lies in the fact that northern knots are entraining and accelerating a layer of protostellar material, a phenomenon undetected in the southern lobe (see Sect.~4.3).  It naturally explains the higher amount of  high-velocity material in the northern lobe and the difference in the mass-loss rates observed between both lobes. The scatter of the knot mass distribution is also larger and we propose it could result from the jet interaction with the ambient medium.\\

\subsection{Ballistic modeling}

The observed variability of the  separation between knots, as discussed in Sect.~\ref{knot-properties}, has usually been interpreted  as the result of the superposition of several ejection modes with different periodicity \citep{Raga2012}, related to velocity or density variations in the ejection process. Here, we propose that the minimum time interval between two consecutive knots corresponds to the period of the variable ejection process and that the radial velocity fluctuations reported in Table~\ref{tab:knots} result from this variability. Then, the apparent bimodal distribution of knot separation can simply be accounted for by dynamical interactions in the nearby environment of protostar A.   

From the knot dynamical ages (see Table~\ref{tab:knots}), we estimate a period $\tau_{ej}= 55\yr$ for the ejection process. It is then possible to translate the knot dynamical age into the number of elapsed periods since its formation, so that each knot can be uniquely identified from its associated period number $n$. This is represented in Fig.~\ref{fig:age-knots}, where the ages of the individual knots are plotted as a function of the number $n$ of periods of the ejection process. The typical uncertainty on the knot location corresponds to $\approx 1\arcsec$, the synthetic beam size (HPFW) of the CO observations, which results into an uncertainty $\Delta n\sim 1$ on the period number $n$. 

Inspection of the knot dynamical ages shows that some events in the northern jet exhibit a counterpart in the southern jet, suggesting that they are associated with the same mass-loss event. This is the case of e.g. knots N0/S1 ($n$= 4), N1/S2 ($n$= 5) and N3/S5 ($n$= 11).  
As discussed in Sect.~5.1, the knots with small $n$ values ($ n < 10$), which are observed at close distance from the protostar, tend to be associated with subsequent ejections, while knots with large period numbers appear to be unevenly distributed (see Fig.~\ref{fig:age-knots}). 

The bottom panel in Fig.~\ref{fig:PV_CO_SO_Main} shows that the young knots N0a-N0b-N0 are ejected  with marked velocity variations. The youngest knot N0a moves at a velocity  $15\kms$ faster than N0b, while they are separated by 1500~au ($\sim 1.8\arcsec$). Hence, it should take about $500\yr$ for N0a to catch up the distance to collide with N0b. The collision is bound to occur at a typical distance of 6500~au, i.e. $8\arcsec$ from the protostar. This collision timescale corresponds to a period number $n\simeq 11$.  As discussed above (see also Fig.~\ref{fig:age-knots}), this value of $n$ is actually the threshold beyond which the knot distribution does not display anymore evidence of subsequent ejections. We also note that on average, the mass of the knots is higher for dynamical timescales longer than $500\yr$. We conclude that the bi-modal knot distribution observed in the jet is most likely the result of collisions between subsequent knots ejected with markedly different velocities. 

We speculate that the elongated appearance of knot S0a might trace the interaction of two subsequent ejections. The spatial structure of knots is out of reach at the angular resolution of the CO data. Observations at higher angular and spectral resolutions are required in order to fully resolve the structure of the knots and to search for the presence of the reverse shocks, which are expected to form between colliding mass-ejecta. 

\citet{Ayala2000} proposed that HH377 traces  the impact of the southern jet in a dense molecular gas clump, explaining that HH377 propagates at a velocity $V \simeq -67\kms$, strongly reduced with respect to the jet velocity itself ($-125\kms$). We propose that the southern knots, once having reached the terminal jet velocity $V\simeq -125\kms$ propagate along the jet until they finally impact the terminal bowshock. In the impact, their velocity suddenly drops while their material "feeds" the HH object.
The distance between S6 and HH377 is  $\approx 4\arcsec$ (3300~au), so that it takes about $270\yr$ for a knot to cover this distance, which is much less than the duration of the mass-loss phase, as estimated from the dynamical age of the oldest  northern knots ($\sim 1500\yr$, see Fig.~\ref{fig:age-knots}). 

The mass of HH377 is $\sim 6\times 10^{-3}\msol$, which amounts to the mass of $\approx$ 6 knots (adopting an average knot mass of $10^{-3}\msol$). This number is in reasonable agreement  with the lack of knots in the range $n$=17--26 in the southern lobe, whereas four knots are detected in the northern jet. 
 From our mass-loss rate analysis, it comes out that the HH377 mass corresponds to the amount of jet material ejected for $\approx 670\yr$. This would imply a dynamical age of $\approx 1410\yr$ for the southern jet, in very good agreement with the dynamical age of knot NB (1440\yr; Table~\ref{tab:knots}), at the tip of the northern jet. 

\section{Jet wiggling}
As recalled in Sect.~1, the cause of the wiggling structure observed in molecular outflows is not completely understood. Several models have been proposed to account for 
the variability of the ejection direction: orbital motion of a binary system \citep{Masciadri-Raga2002}, jet precession due to tidal interactions between the associated protostellar disk and a non-coplanar binary companion \citep{Terquem1999}, misalignment between the disk rotation axis and the ejection mechanism \citep{Frank2014}. The latter model requires only a single protostar. 

The limited angular resolution of our observations (320--800 au) does not allow to probe the disk scale nor the jet launch region. Here, we consider the two following classes of periodic mass-ejection models proposed to account for the wiggling pattern of protostellar outflows: i)~jet precession; ii)~orbital motion around a binary companion. Both models produce a garden-hose effect \citep{Raga-Canto-Biro1993}: close to the source the jet propagates almost parallel to the main jet axis while further downstream, the side-to-side wiggling is amplified as the fluid parcels ejected in different directions diverge from each other.  Outflow morphology allows discriminating  between both models: an anti-symmetric {\it S shape} is observed in the precession scenario, while a mirror symmetric {\it W shape} is observed in the orbital motion scenario \citep[e.g.,][]{Raga-Canto-Biro1993, Masciadri-Raga2002, Noriega-Crespo2011, Velazquez2013, Hara2021}. 

\subsection{Precession}
\begin{figure}
\centering
\includegraphics[width=0.45\textwidth]{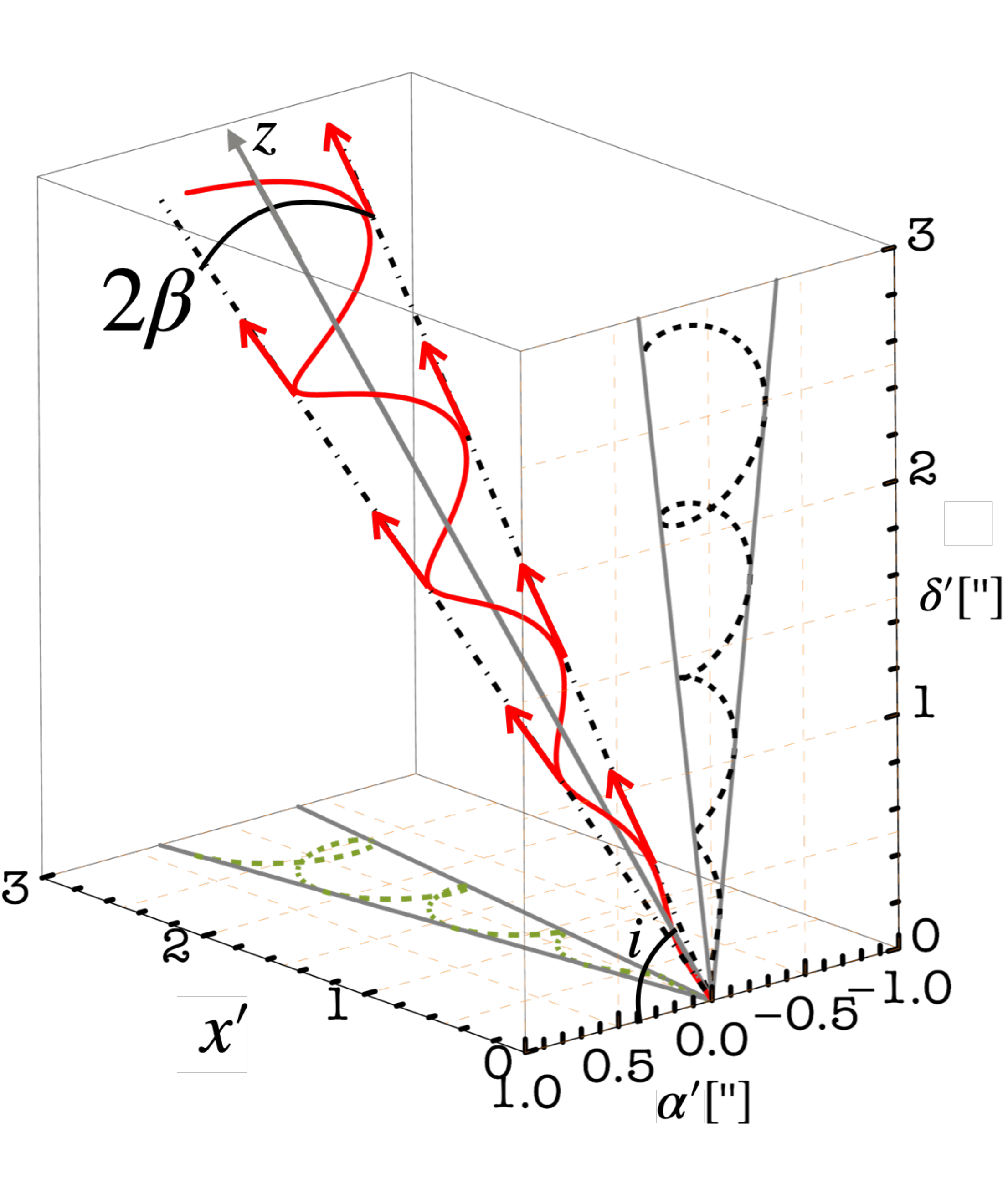}
\includegraphics[width=0.3\textwidth]{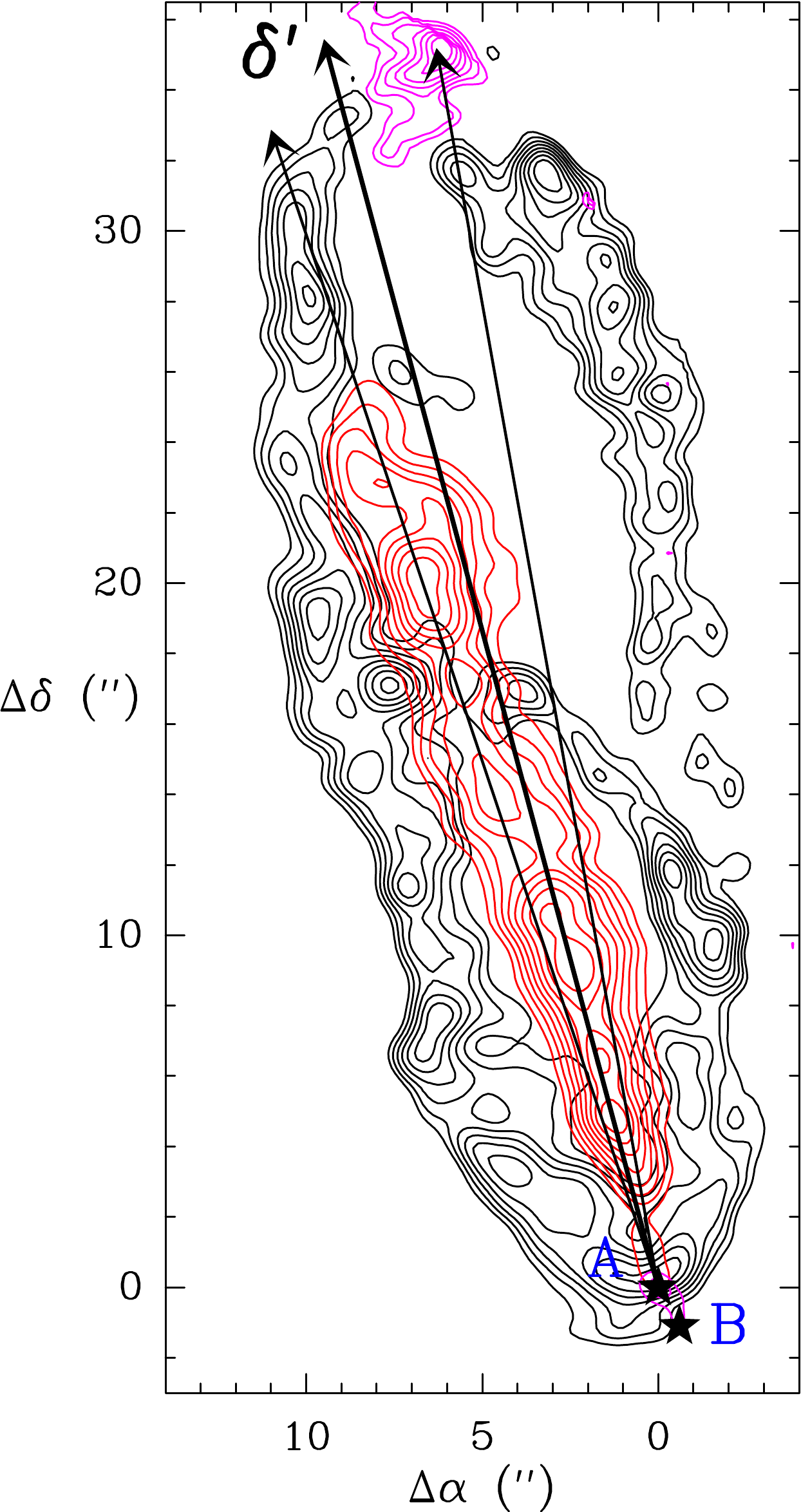}
    \caption{Jet precession in Cep\,E. (top)~Wiggling model. The jet precession axis is represented by the continuous gray line. The jet moves over a cone of half-opening angle $\beta$. It makes an inclination angle $i$ with respect to the plane of the sky ($\alpha^{\prime}$,$\delta^{\prime}$). The locus of the ejection direction is drawn by the red curve. In the plane of the sky ($\alpha^{\prime}$,$\delta^{\prime}$), the projection of the locus is represented by the black line. (bottom)~Projected view of the jet axis associated with the current high-velocity emission, the previous ejection associated with knot NB and the precession axis $\delta^{\prime}$, superimposed on the CO 2--1 outflow emission (see Fig.~1). }
    \label{fig:jetModel}
\end{figure}

We adopt a more detailed modeling of the jet precession than the previous works by \citet{Eisloeffel1996} and \citet{Noriega-Crespo2014} by making use of the spatial distribution of knots  to constrain the dynamical parameters of the ejection process, under the assumption of ballistic motion. We consider a jet precessing  inside a cone of main axis $z$ and half opening angle $\beta$ (see top panel of Fig.~\ref{fig:jetModel}). The jet makes a spiral pattern described by the following parametric equations \citep{Raga-Canto-Biro1993}:
\begin{equation}
     x= z \tan \beta \sin\left(\frac{2\pi}{\lambda}z+\phi\right)\label{eq:x}
\end{equation}
and
\begin{equation}
   y= z \tan \beta \cos\left(\frac{2\pi}{\lambda}z+\phi\right),\label{eq:y}
\end{equation}
where $\lambda$ is the spatial period and $\phi$ is the phase angle.
In general, the precession axis makes an inclination angle $i$ with respect to the plane of the sky and the projected coordinates ($\alpha'$,$\delta'$) in this plane are given by
\begin{equation}
    (x',\alpha',\delta')=(x\cos i+ z\sin i, y,  z\cos i-x\sin i). \label{eq:projected}
\end{equation}
We adopted for Cep\,E an inclination angle $i$= $47^\circ$, as mentioned in Sect.~\ref{sec:source}.  We first determined the parallactic angle of the jet precession axis $z$ in the plane of the sky, from the average position angle of the lines formed by the source and the knots, including NB and HH377. Proceeding independently for the northern and southern lobes, we find  a misalignment of about $7^\circ$ between both axes. The configuration in the Northern outflow is summarized in the bottom panel of Fig.~\ref{fig:jetModel}: the loci of the current jet, the precession axis $\delta'$ and the direction of the previous ejection associated with knot NB. 

\begin{figure}[!ht]
\centering
 \includegraphics[width=0.48\textwidth]{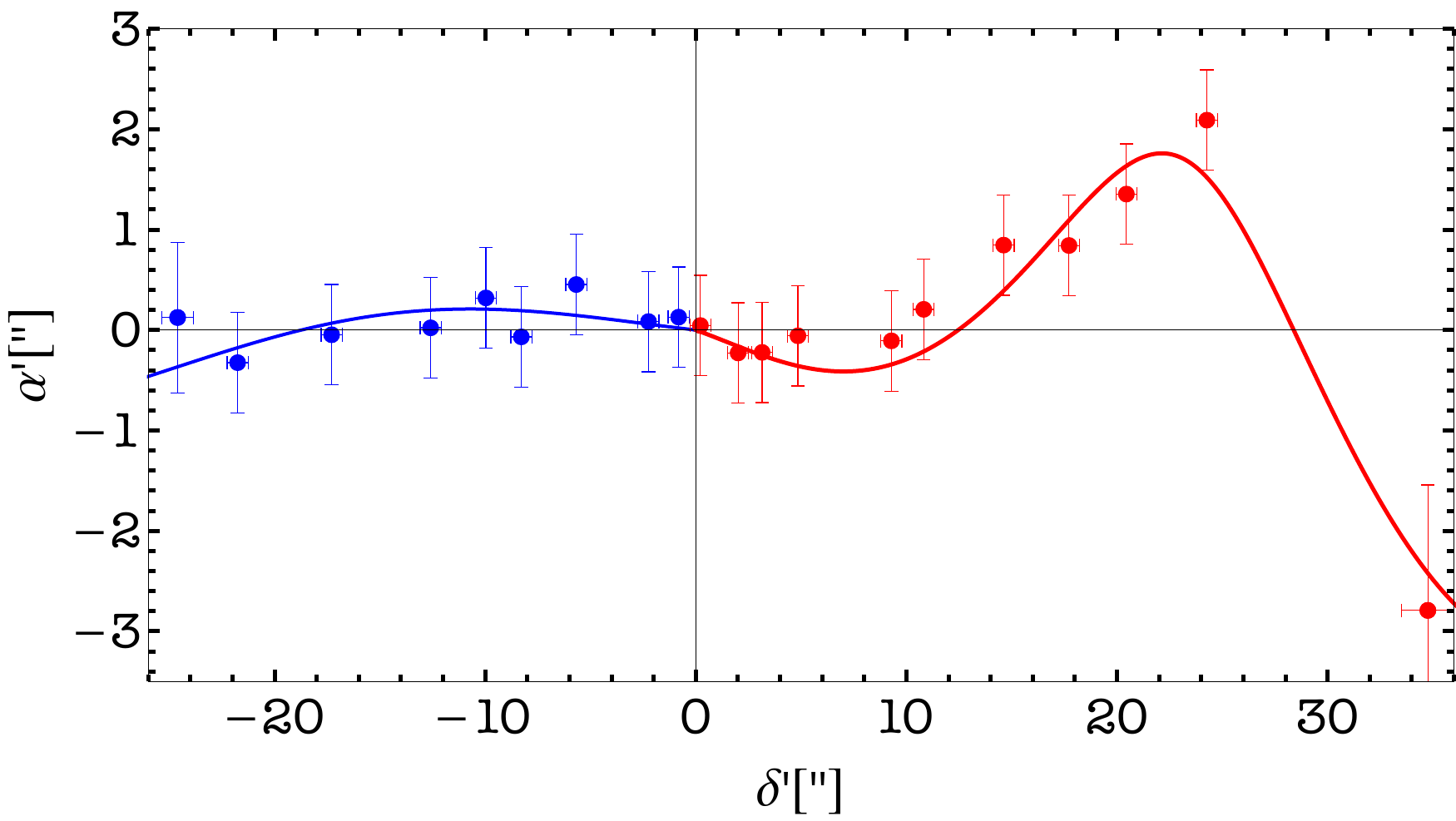}
\caption{Positions of the CO knots with respect to the precession axis in the northern (red) and southern (blue) lobe. The solid line draws the  best fitting precession model to the observational data.}
    \label{fig:precession}
\end{figure}

After determining the orientation of the jet precession axis $z$, we have computed the coordinates ($\alpha'$,$\delta'$) of each knot in the plane of the sky, which are reported in Fig.~\ref{fig:precession}. 
The northern (red) knots show an increasing deviation in the positive $\alpha^\prime$ direction until $\delta'$= $26\arcsec$. Further away, NB is  detected at negative offsets $\alpha'$= $-3\arcsec$ from the precession axis. 
In the southern lobe, (blue) knots appear to lie along the precession axis, with small positive offsets at less than 8000 au from the protostar.  The knot spatial distribution displayed in Fig.~\ref{fig:precession} around the driving protostar at $\delta$= $0\arcsec$  is consistent with an S-shape symmetry. A W-shape symmetry would require a knot distribution with negative offsets in the Southern lobe, which is not compatible with the observational data. We conclude that comparison of the northern and southern knot spatial distributions unambiguouly favors a S-shape geometry.  

Using Eqs.~\ref{eq:x}--\ref{eq:projected}, we have then searched for the parameter set ($\lambda$, $\beta$, $\phi$) which best reproduces the location of the knots in the plane of the sky. Note that we modeled the northern and southern lobes separately. From the total jet velocity $V_t$, we derive the precession period $\tau_p$= $\lambda/V_t$. The best fitting model was obtained through a least-squares method and is drawn as a (red/blue) continuous curve  in Fig.~\ref{fig:precession}. The  parameters of the best fitting model are presented in Table~\ref{tab:prec}. 

Our model succeeds in reproducing the knot spatial distributions both in the northern  and the southern jet.  In the latter case, we note however that the variations in the knot distribution are  smaller with respect to the position uncertainties, hence enlarging uncertainties in the parameters of the best fitting solution (see Table~\ref{tab:prec}).  
The values of the phase angle $\phi$ and the precession period $\tau_P$ are found similar between the northern and southern lobes. This is consistent with symmetric ejections, in agreement with the observational data. 

As can be seen in Table~\ref{tab:prec}, differences are observed  between the northern and southern lobes in the alignment of the precession axes, the jet velocities and the half-opening precession angle. We propose that this is  related to the different environmental conditions between the outflow lobes \citep{Hirth1994,Velazquez2014}.

\begin{table}[ht]
\begin{center}
\caption{Parameters of the jet precession model}
{\small
\begin{tabular}{lccccc}
\hline \hline
 &  $\lambda$ & $V_t$\tablefootmark{a}& $\beta$ &  $\tau_p$ &$\phi$ \\ 
 &(\arcsec)& (\kms) & ($^\circ$) & ($yr$) &($^{\circ}$)\\\hline
North & $53.3\pm 0.5$ & $106$ & $3.3 \pm 0.2$ & $1960\pm 20$ & $154 \pm 5$ \\
South & $76.7\pm 4.5$ & 163 & $1.2\pm 0.2$ & $1840\pm 120$ & $168\pm 10$\\
\hline
\label{tab:prec}
\end{tabular}
}
\end{center}
\tablefoot{Length period $\lambda$, half-opening angle $\beta$, total de-projected jet velocity $V_t={V_j}/{\sin{i}}$ (with $i=47^\circ$), precession time period $\tau_p$ used in the precession model for each outflow lobe and the phase angle $\phi$.}
\tablefoottext{a}{Measured from the CO observations.}
\end{table}

Precession changes the direction of the jet propagation with respect to the plane of the sky, and therefore affects the knot radial velocity. Taking into account the inclination of the Cep\,E jet with respect to the plane of the sky, the half-opening angle of its precession cone can yield velocity variations up to $15\kms$.
Hence, the combined effect of the ejection variability and precession is expected to generate velocity variations of $25 \kms$ or more, which could explain the higher radial velocity reported for NB. To conclude, velocity variations as large as those measured between NB and younger knots are consistent with the wiggling dynamics of the Cep\,E jet; such variations are expected to be rare as the jet dynamical age is similar to the precession period. 

\subsection{Origin of the jet wiggling}
\citet{Masciadri-Raga2002} first proposed that the orbital motion of a binary system could account for the wiggling morphology of outflows. The authors found that the same description of a precessing jet model could be applied at a distance from the precession axis larger than the binary separation $r_0$. In this case, the half-opening angle is determined by the ratio between the orbital velocity and the total jet velocity $\tan \beta= V_o/V_t$. Also, the authors related the distance $\Delta z$  between a maximum and a minimum in $y$ ($\Delta z\sim 23\arcsec$ for Cep\,E), to the separation $r_0$ of the two binary components:
\begin{equation}
    \Delta z= \frac{2 \pi r_0}{\tan \beta},
\end{equation}
which implies $r_0 \sim 140 $~au and an orbital period $\tau_o$ of 
\begin{equation}
    \tau_0=\frac{2\pi r_0}{V_o} =\frac{\Delta z}{V_t} \sim 700 \rm{yr},
\end{equation}
at the distance of Cep\,E-mm ($819\pc$).  Then, we estimate a typical mass 
\begin{equation}
\label{eq:kepler}
    M_s=\frac{16 \pi^2 r_0^3}{\tau_o^2 G} \sim 20 \rm{M}_\odot 
\end{equation}
for the components of the binary. Note that in their original work \cite{Masciadri-Raga2002} derived this relation under the assumption of equal mass components. 
This value is too high to be reconciled with the luminosity of the protostellar core ($100\lsol$) and the core masses derived from the dust thermal emission at arcsec scale by \cite{Ospina-Zamudio2018}. Applying Eq.~\ref{eq:kepler} to protostar B, which lies at a distance of $1.2\arcsec$ ($\sim 9000$ au) from A and assuming the same orbital period the mass required would be even larger than $\sim 20M_\odot$. 

As discussed in the previous section, the knot spatial distribution along the jet  (Fig.~\ref{fig:precession}) appears in better agreement with an S-shape than a W-shape symmetry. 
The symmetry of the distribution therefore supports jet precession rather than an orbital motion scenario as origin of the outflow direction change in Cep\,E.  

To summarize, the orbital motion scenario fails to account for the properties of the 
knot spatial distribution in the Cep\,E protostellar jet. Our analysis favors a jet precession scenario as origin of the outflow wiggles. As mentioned above in this Section, \citet{Terquem1999} proposed that jet precession originates in the tidal interaction between the protostellar disk driving the jet and a non-coplanar binary companion. The authors applied their analytic model to the case of Cep\,E based on the parameters estimated by \citet{Eisloeffel1996}. They concluded that the system is a very close binary with a distance  in the range 4 -- 20~au and a small disk radius R= 1 -- 10~au. Taking into account the much longer precession timescale implied by our new model, $2000\yr$ instead of $400\yr$, the separation of the binary increases by a factor $\sim 2$ up to  8 -- 40~au (0\farcs01 -- 0\farcs05) but unfortunately the linear scales involved remain out of reach to the NOEMA interferometer. 
We note that also in the scenario proposed by \citet{Terquem1999}, protostar B cannot be responsible for the precession observed in the jet from protostar A. 

\section{Conclusions}
We have carried out a detailed kinematical study of the high-velocity jet of the intermediate-mass Class 0 protostar Cep\,E-A using observations from the IRAM interferometer at (sub)arcsec resolution in the CO 2--1 and SO $N_J$= $5_4$--$4_3$ lines. Our conclusions are as follows:
\begin{itemize}

\item~The CO high-velocity jet consists of two components: a central component of diameter $\leq 400$~au associated with high-velocity molecular knots (also detected in SO) and a surrounding gas layer  accelerated on a length scale $\delta_0\approx$ 700~au, whose diameter gradually increases up to several 1000~au at about 2000~au from the protostar. 

\item~Along the main jet axis, the CO gas velocity  can be simply fitted as a function of distance $\delta$ to the protostar by the  relation $(V_r-V_{lsr})/V_0$= $\exp(-\delta_0/\delta)$
where $V_{lsr}$= $-10.9\kms$ is the ambient cloud velocity, $V_0$ is the terminal velocity in the jet lobe relative to the source $\approx +75\kms$ ($-118\kms$) and $\delta_0$ is the length scale $\approx 700$~au (600~au) in the northern (southern) hemisphere. 
Analysis of the CO Position-Velocity diagrams shows that the jet layer consists of quiescent protostellar material entrained by extremely-high velocity knots. The material entrainment appears to be more efficient in the northern than in the southern lobe. 

\item~The jet molecular gas emission reveals the presence of 18 bright knots in either lobes of the jet with a dynamical timescale of $55\yr$ between subsequent ejections close ($< 1000$~au) to the protostar.  The knots have a typical size of $2\arcsec$, with a tendency to increase with  distance to the protostar. We have identified  knots NB and HH377 as the terminal bowshocks of the northern and southern lobes, respectively.

\item~Assuming a standard CO abundance of $10^{-4}$ relative to $\htwo$, a simple radiative transfer analysis in the LVG approximation with MadeX yields a typical mass of $10^{-3}\msol$ for the bullets. SO observations at $0.4\arcsec$ show that the knots leave the inner protostellar regions ($\sim 200$~au) at high-velocity, with marked velocity fluctuations between subsequent ejections. The cumulative knot mass function allows to estimate mass-loss rates of $2.7\times10^{-5}\msol\yrmu$ and $8.9\times 10^{-6}\msol\yrmu$ in the northern and southern lobes, respectively. The jet mass-loss rates appear steady with time. The higher mass-loss rate value in the North is consistent with protostellar material entrainment by the jet. 

\item~ The knot interval distribution between subsequent knots is approximately bi-modal with intervals of $\sim 55$ yr and $\sim 150$ at short ($< 12\arcsec$) and large ($> 12\arcsec$) distance from the protostar, respectively. In the northern jet, the velocity fluctuations are large enough ($\sim 15\kms$) that the spatial separation between knots can be overcome in a dynamical timescale of $\simeq 500\yr$, when knots will have travelled a typical distance of $12\arcsec$ ($10^4$~au) from the protostar. This naturally explains lack of knots at large distances from the source. This result is also supported by the higher masses found for the knots at larger distances. 

\item Analysis of the knot spatial distribution suggests precession as origin of the wiggling structure of the outflow. Applying the Masciadri-Raga formalism, we obtain an accurate description of the jet morphology, taking into account the outflow inclination angle with respect to the plane of the sky. The same model allows us to discard binary orbital motions as the source of the Cep\,E-mm jet wiggling, since it would require protostellar masses too large to be reconciled with the source luminosity and mass. We propose that jet precession takes its origin in tidal perturbations of the associated protostellar disk by a binary companion, yet to be identified, in agreement with previous work by \cite{Terquem1999}.

\end{itemize}

\vspace{5mm}

\begin{acknowledgements}
Based on observations carried out with the IRAM NOEMA interferometer. IRAM is supported by INSU/CNRS (France), MPG (Germany) and IGN (Spain).
AS, BL, PR-RO, acknowledge support from the European Union’s Horizon 2020 research and innovation program under the Marie Skłodowska-Curie grant agreement No 811312 for the project "Astro-Chemical Origins” (ACO) and the  European Research  Council (ERC) under the  European  Union’s Horizon 2020 research and innovation program for the Project “The Dawn of Organic Chemistry” (DOC) grant agreement No 741002. DSC is supported by an NSF Astronomy and Astrophysics Postdoctoral Fellowship under award AST-2102405. A.V. and A.P. are the members of the Max Planck Partner Group at the Ural Federal University. A.P. and A.V. acknowledge the support of the Russian Ministry of Science and Education via the State Assignment Contract no. FEUZ-2020-0038.
\end{acknowledgements}


\begin{thebibliography}{}

\bibitem[\protect\citeauthoryear{Ayala et al.}{2000}]{Ayala2000} 
Ayala S., Noriega-Crespo A., Garnavich P.~M., et al., 2000, AJ, 120, 909

\bibitem[\protect\citeauthoryear{Bachiller}{1996}]{Bachiller1996}
Bachiller, R., 1996, ARA\&A, 34, 111

\bibitem[\protect\citeauthoryear{Bally}{2016}]{Bally2016} 
Bally J., 2016, ARA\&A, 54, 491

\bibitem[\protect\citeauthoryear{Cabrit et al.}{1997}]{Cabrit1997}
Cabrit,S., Raga, A., Gueth, F., in Herbig-Haro Flows and the Birth of Stars; IAU Symposium No. 182, Edited by Bo Reipurth and Claude Bertout. Kluwer Academic Publishers, 1997, p. 163-180

\bibitem[\protect\citeauthoryear{Ceccarelli et al.}{2017}]{Ceccarelli2017} 
Ceccarelli C., Caselli P., Fontani F., et al., 2017, ApJ, 850, 176

\bibitem[\protect\citeauthoryear{Cernicharo}{2012}]{Cernicharo2012} 
Cernicharo J., 2012, EAS, 58, 251

\bibitem[\protect\citeauthoryear{Chini et al.}{2001}]{Chini2001} 
Chini R., Ward-Thompson D., Kirk J.~M., et al., 2001, A\&A, 369, 155

\bibitem[\protect\citeauthoryear{Codella et al.}{2007}]{Codella2007}
Codella, C., Cabrit, S., Gueth, F., et al., 2007, A\&A, 462, L53

\bibitem[\protect\citeauthoryear{Codella et al.}{2014}]{Codella2014} Codella C., Cabrit S., Gueth F., et al., 2014, A\&A, 568, L5. 

\bibitem[\protect\citeauthoryear{Crimier et al.}{2010}]{Crimier2010} Crimier N., Ceccarelli C., Alonso-Albi T.,  et al., 2010, A\&A, 516, A102. 

\bibitem[\protect\citeauthoryear{Eisloeffel et al.}{1996}]{Eisloeffel1996} 
Eisloeffel J., Smith M.~D., Davis C.~J., Ray T.~P., 1996, AJ, 112, 2086. doi:10.1086/118165

\bibitem[\protect\citeauthoryear{Ferrero, G{\'o}mez, \& Gunthard}{2015}]{Ferrero2015a} 
Ferrero L.~V., G{\'o}mez M., Gunthard G., 2015, BAAA, 57, 126

\bibitem[\protect\citeauthoryear{Frank et al.}{2014}]{Frank2014} 
Frank, A., Ray, T., Cabrit, S., et al., 2014, Protostars and Planets VI, eds
H.~Beuther, R.S.~Klessen, C.P.~Dullemond \& T.~Henning (Tucson, AZ: University of Arizona Press), 
451

\bibitem[\protect\citeauthoryear{G\'omez-Ruiz et al.}{2012}]{Gomez-Ruiz2012}
G\'omez-Ruiz, A. I., Gusdorf, A., Leurini, S., et al., 2012, A\&A, 542, L9

\bibitem[\protect\citeauthoryear{Gueth, Guilloteau, \& Bachiller}{1996}]{Gueth1996}
Gueth, F., Guilloteau, S., \& Bachiller, R. 1996, A\&A, 307, 891

\bibitem[\protect\citeauthoryear{Gusdorf et al.}{2017}]{Gusdorf2017} 
Gusdorf A., Anderl S., Lefloch B., et al., 2017, A\&A, 602, 8

\bibitem[\protect\citeauthoryear{Hara et al.}{2021}]{Hara2021} 
Hara C., Kawabe R., Nakamura F.,  et al., 2021, ApJ, 912, 34. 

\bibitem[\protect\citeauthoryear{Hirth et al.}{1994}]{Hirth1994} Hirth G.~A., Mundt R., Solf J., Ray T.~P., 1994, ApJL, 427, L99. %doi:10.1086/187374

\bibitem[\protect\citeauthoryear{Lefèvre et al.}{2017}]{Lefevre2017}
Lefèvre, C., Cabrit, S., Maury, A.J., et al., 2017, A\&A, 604, L1.

\bibitem[\protect\citeauthoryear{Karnath et al.}{2019}]{Karnath2019} 
Karnath N., Prchlik J.~J., Gutermuth R.~A., et al., 2019, ApJ, 871, 46.

\bibitem[\protect\citeauthoryear{Lefloch, Eisloeffel \& Lazareff}{1996}]{Lefloch1996} 
Lefloch B., Eisloeffel J., Lazareff B., 1996, A\&A, 313, L17

\bibitem[\protect\citeauthoryear{Lefloch et al.}{2002}]{Lefloch2002} 
Lefloch B., Cernicharo J., Rodr{\'\i}guez L.~F., et al., 2002, ApJ, 581, 335

\bibitem[\protect\citeauthoryear{Lefloch et al.}{2007}]{Lefloch2007}
Lefloch, B., Cernicharo, J., Reipurth, B., et al., 2007, ApJ, 658, 498

\bibitem[\protect\citeauthoryear{Lefloch et al.}{2015}]{Lefloch2015}
Lefloch, B., Gusdorf, A., Codella, C., et al. 2015, A\&A, 581, A4

\bibitem[\protect\citeauthoryear{Maret et al.}{2009}]{Maret2009} 
Maret S., Bergin E.~A., Neufeld D.~A., et al., 2009, ApJ, 698, 1244. 

\bibitem[\protect\citeauthoryear{Masciadri \& Raga}{2002}]{Masciadri-Raga2002}
Masciadri, E., Raga, A. C., 2002, ApJ, 568, 733 

\bibitem[\protect\citeauthoryear{Nony et al.}{2020}]{Nony2020}
Nony, T., Motte, F., Louvet, F., et al., 2020, 636, A\&A, 38

\bibitem[\protect\citeauthoryear{Noriega-Crespo et al.}{2011}]{Noriega-Crespo2011} Noriega-Crespo A., Raga A.~C., Lora V., Stapelfeldt K.~R., Carey S.~J., 2011, ApJL, 732, L16. %doi:10.1088/2041-8205/732/1/L16

\bibitem[\protect\citeauthoryear{Noriega-Crespo et al.}{2014}]{Noriega-Crespo2014} 
Noriega-Crespo A., Raga A.~C., Moro-Mart{\'\i}n A., et al., 2014, NJPh, 16, 105008. 

\bibitem[\protect\citeauthoryear{Ospina-Zamudio et al.}{2018}]{Ospina-Zamudio2018} Ospina-Zamudio, J., Lefloch, B., Ceccarelli, C., et al., 2018, A\&A, 618, 145. 

\bibitem[\protect\citeauthoryear{Ospina-Zamudio et al.}{2019}]{Ospina-Zamudio2019} Ospina-Zamudio J., Lefloch B., Favre C., et al., 2019, MNRAS, 490, 2679

\bibitem[\protect\citeauthoryear{Plunkett et al.}{2015}]{Plunkett2015} 
Plunkett A.~L., Arce H.~G., Mardones D., et al., 2015, Nature, 527, 70

\bibitem[\protect\citeauthoryear{Podio et al.}{2016}]{Podio2016}
Podio, L., Codella, C., Gueth, F., et al. 2016, A\&A, 593, L4

\bibitem[\protect\citeauthoryear{Podio et al.}{2021}]{Podio2021}
Podio, L., Tabone, B., Codella, C., et al., 2021, A\&A, 648, 45

\bibitem[\protect\citeauthoryear{Raga et al.}{1990}]{Raga1990} 
Raga, A.C., Cantó, J., Binette, Calvet, N., 1990, ApJ, 364, 601 

\bibitem[\protect\citeauthoryear{Raga \& Cabrit}{1993}]{Raga-Cabrit1993} 
Raga, A.C., Cabrit S., 1993, A\&A, 278, 267

\bibitem[\protect\citeauthoryear{Raga, Cantó \& Biro}{1993}]{Raga-Canto-Biro1993}
Raga, A.C., Cantó, J., \& Biro, S., 1993, \mnras, 260, 163 

\bibitem[\protect\citeauthoryear{Raga et al.}{2002}]{Raga2002} 
Raga, A.C., Velázquez, P.F., Cantó, J., Masciadri, E., 2002, ApJ, 395, 647 

\bibitem[\protect\citeauthoryear{Raga et al.}{2012}]{Raga2012} 
Raga, A.C., Rodr{\'\i}guez-Gonz{\'a}lez A., Noriega-Crespo A., Esquivel A., 2012, ApJL, 744, L12. 

\bibitem[\protect\citeauthoryear{Reipurth \& Bally}{2001}]{Reipurth-Bally-2001} 
Reipurth,B., Bally, J., 2001, ARA\&A, 39, 403

\bibitem[\protect\citeauthoryear{Tafalla et al.}{2010}]{Tafalla2010} 
Tafalla M., Santiago-Garc{\'\i}a J., Hacar A., Bachiller R., 2010, A\&A, 522, A91. 

\bibitem[\protect\citeauthoryear{Terquem et al.}{1999}]{Terquem1999} 
Terquem, C., Eislöffel, J., Papaloizou, C.B., Nelson, R.P., 1999, ApJ, 512, L131 

\bibitem[\protect\citeauthoryear{Tychoniec et al.}{2019}]{Tychoniec2019} 
Tychoniec, {\L}., Hull C.L.H., Kristensen L.E., et al., 2019, A\&A, 632, A101

\bibitem[\protect\citeauthoryear{Vel{\'a}zquez et al.}{2013}]{Velazquez2013} Vel{\'a}zquez P.~F., Raga A.~C., Cant{\'o} J., et al. , 2013, MNRAS, 428, 1587. %doi:10.1093/mnras/sts139

\bibitem[\protect\citeauthoryear{Vel{\'a}zquez et al.}{2014}]{Velazquez2014} Vel{\'a}zquez P.~F., Riera A., Raga A.~C., Toledo-Roy J.~C., 2014, ApJ, 794, 128. %doi:10.1088/0004-637X/794/2/128

\bibitem[\protect\citeauthoryear{Wilkin}{1996}]{Wilkin1996}
Wilkin,F.P., 1996, ApJ, 459, L31

\bibitem[\protect\citeauthoryear{Yang et al.}{2010}]{Yang2010}
Yang B., Stancil P.~C., Balakrishnan N., Forrey R.~C., 2010, ApJ, 718, 1062


\end{thebibliography}
\end{document}